\documentclass[a4paper,11pt]{article}
\pdfoutput=1 

\usepackage{jcappub} 
\usepackage[T1]{fontenc} 

\usepackage{comment}
\usepackage{amsmath}
\usepackage{amssymb}
\usepackage[scr=boondox]{mathalpha}
\usepackage{multirow}
\usepackage{xcolor}
\usepackage{hyperref}
\usepackage[normalem]{ulem}

\newcommand{\dd}{{\rm d}}
\newcommand{\B}{{\rm B}}
\newcommand{\X}{{\rm X}}
\newcommand{\e}{{\rm e}}

\DeclareMathAlphabet\mathbfcal{OMS}{cmsy}{b}{n}

\newcommand{\repl}[1]{#1}

\newcommand{\sig}{\Sigma}

\subheader{ULB-TH/26-18}

\title{
Reionization, UV Luminosity and 21$\,$cm Sensitivity to 
Primordial Magnetic Fields: Impact of Energy Losses}
\author[a]{G.~Facchinetti,}
\author[a, b]{A.~Korochkin,}
\author[a, c]{L.~Lopez-Honorez,}
\author[a]{J.~R.~Schwagereit}
\date{April 2026}

\affiliation[a]{Service de Physique Th\'{e}orique \& Brussels Laboratory of the Universe BLU-ULB, C.P. 225, Universit\'{e} Libre de Bruxelles,  Boulevard du Triomphe, B-1050 Brussels, Belgium}

\affiliation[b]{Institute for Nuclear Research of the Russian Academy of Sciences, Moscow, 117312 Russia}

\affiliation[c]{Theoretische Natuurkunde \& The International Solvay Institutes, \\
Vrije Universiteit Brussel, Pleinlaan 2, B-1050 Brussels, Belgium}

\emailAdd{gaetan@facchinetti-corne.fr}
\emailAdd{aa.korochkin@physics.msu.ru}
\emailAdd{justus.roman.schwagereit@ulb.be}
\emailAdd{laura.lopez.honorez@ulb.be}

\abstract{
Magnetic fields  with field strengths between $10^{-17}\,$G  and a few Nanogauss are expected to exist today in the intergalactic medium (IGM). Their origin is unknown, but may be of primordial nature, in which case they would have influenced the thermal and ionization history of the IGM as well as the growth of small-scale matter perturbations. In this work, we revisit constraints on Primordial Magnetic fields (PMFs) by consistently accounting for their energy losses through ambipolar diffusion and decaying turbulences from recombination through the epoch of reionization, which progressively reduces the magnetic field strength over time. We implement these effects in { \tt HyRec}  and {\tt exo21cmFAST} to model the interplay between PMFs and astrophysical processes up to reionization. Using a neural-network emulator ({\tt NNERO}), we perform a MCMC analysis that combines late-time probes of the reionization history and galaxy UV luminosity functions. We find that including PMF energy losses significantly relaxes previous bounds, as the reduced field strength suppresses  their imprint on observables. Employing a Fisher matrix analysis, we estimate the sensitivity of the 21$\,$cm signal experiment HERA to the PMFs' imprint on intergalactic medium perturbations and show that 21$\,$cm cosmology can significantly improve on current bounds\repl{, depending on assumptions on the astrophysics}.  Our results highlight the importance of modeling PMF evolution self-consistently with the IGM evolution to extract current bounds and future sensitivities.

}


\arxivnumber{2604.22703}

\begin{document}

\maketitle

\section{Introduction}
\label{intro}

The existence of non-zero magnetic fields in the voids of the universe's large-scale structure is suggested by the non-observation of secondary $\gamma$-ray emissions from distant active galactic nuclei~\cite{2010Sci...328...73N} and gamma-ray bursts~\cite{vovk_grb}. While the mechanism responsible for generating these fields remains unknown, one hypothesis suggests that they were generated and amplified during the epoch of structure formation and subsequently transported into voids via magnetized outflows from galaxies or AGN jets~\cite{Bertone:2006mr}. However, the most recent models indicate that galactic outflows may lack the required power to produce such volume-filling magnetic fields in voids~\cite{2018MNRAS.480.5113M, 2022A&A...660A..80B}.

Alternatively, magnetic fields in voids could be of primordial (cosmological) origin. For example, primordial magnetic fields (PMFs) could have been generated during inflation through a coupling between the electromagnetic and scalar inflaton fields~\cite{1988PhRvD..37.2743T, 1992ApJ...391L...1R, Subramanian:2015lua}. In this scenario, the coherence length of such PMFs could reach Mpc scales, see e.g.~\cite{Sharma:2017eps}. PMFs could also have been produced during cosmic electroweak or quantum chromodynamics phase transitions~\cite{1983PhRvL..51.1488H, Vachaspati:2020blt, Neronov:2020qrl}. In that case, their initial coherence scale  would be smaller than the Hubble horizon at the time of the transition but could later increase to tens or hundreds of kiloparsecs by the end of recombination due to magnetohydrodynamical processes.

Observations in the local universe (at $z \sim 1$) constrain the intergalactic magnetic field (IGMF) strength to the range between approximately $10^{-17}\,$G up to $10^{-9}\,$G. The lower limits are derived from $\gamma$-ray observations of distant sources~\cite{2010Sci...328...73N, MAGIC:2022piy, Blunier:2025ddu, Vovk:2025rkr}, while the upper limits come from the absence of a redshift evolution of rotation measures of extragalactic sources~\cite{Blasi:1999hu, Pshirkov:2015tua, Neronov:2024qtk}. 
If the IGMF is of primordial origin, i.e. produced by some process in the early universe, it could also affect processes at higher redshifts, such as during cosmic recombination and reionization, see e.g.~\cite{Durrer:2013pga}. PMFs can therefore be constrained from CMB observations~\cite{ Planck:2015fie, Chluba:2015lpa,  Paoletti:2018uic, Cruz:2023rmo,Zucca:2016iur, Shaw:2009nf, durrer1998microwave, Jedamzik:1999bm, Zizzo:2005az, Kunze:2013uja, Saga:2017wwr, Barrow:1997mj, Jedamzik:2018itu, Subramanian:1998fn,2002MNRAS.335L..57S,Mack:2001gc,Lewis:2004kg,Kahniashvili:2005xe,Chen:2004nf,Lewis:2004ef,Tashiro:2005hc, Yamazaki:2006bq,Giovannini_2006,Paoletti:2010rx,2012PhRvD..86d3510S, Kunze:2013iwa,Kunze:2011bp,Kunze:2010ys,Paoletti:2012bb,Ade:2015xua,Sethi:2004pe, Seshadri:2000ky,Paoletti:2022gsn,Schleicher:2011jj,Pandey:2014vga,Jedamzik:2023csc,Jedamzik:2023rfd}, from Lyman-$\alpha$ forest observations~\cite{Pandey:2012ss, Pavicevic:2025gqi, Garcia-Gallego:2026phh}, from line intensity mapping~\cite{Adi:2023qdf}, and reionization and 21$\,$cm cosmology~\cite{Pandey:2014vga, Bera:2020jsg, Katz:2021iou,Minoda:2022nso, Cruz:2023rmo,Bhaumik:2024efz,Novosyadlyj:2025zxs,Bhaumik:2026mlc}.
The presence of PMFs would also have impacted galaxy and star formation processes~\cite{1978ApJ...224..337W, Shibusawa:2014fva, Schleicher:2009zb, Sanati:2020oay, ralegankar2024primordial, Marinacci:2015ria,Katz:2021iou}.

In this paper we focus on the evolution of the PMFs after epoch of recombination. Specifically, we derive the time evolution of the IGM's ionized hydrogen fraction and temperature together with the  magnetic field energy density evolution, $\rho_\B({\bf x},z)={\bf B}({\bf x},z)^2 / (2\mu_0)$, where ${\bf x}$ denotes comoving  position and $z$ the redshift, while accounting for PMF energy injection into the plasma. In past literature,  it has mainly been assumed that   $ \rho_\B({\bf x},z)\propto (1+z)^4$, which would be the case when neglecting energy losses through  IGM heating. In this paper, we   track the PMF energy density decay due to ambipolar diffusion and decaying turbulence effects. For that purpose, those effects have consistently been implemented into the public codes {\tt HyRec-2}\footnote{\href{https://github.com/gaetanfacchinetti/HYREC-2}{https://github.com/gaetanfacchinetti/HYREC-2}} (based on  {\tt HyRec}~\cite{Ali-Haimoud:2010hou, Lee:2020obi})  and {\tt exo21cmFAST}\footnote{\href{https://github.com/gaetanfacchinetti/exo21cmFAST/tree/pmf}{https://github.com/gaetanfacchinetti/exo21cmFAST (branch pmf)}}~\cite{Facchinetti:2023slb} (based on {\tt 21cmFAST}~\cite{Mesinger:2010ne}). Furthermore, we also account for the PMF-induced boost~\cite{Kim:1994zh, yamazaki2006effect, barrow2007cosmology, 2012PhRvD..86d3510S, 2012JCAP...11..055F, ralegankar2024primordial, Ralegankar:2024arh, Subramanian:1997gi,  Jedamzik:1996wp} to small-scale structure formation due to Lorentz pressure, following the recent works of~\cite{Adi:2023doe,Adi:2023qdf} to re-evaluate this effect.
 
We self-consistently use these inputs to revisit the imprint of PMFs on reionization, galaxy UV luminosity functions (UV LFs), and the 21$\,$cm signal, see e.g.~\cite{Sethi:2004pe,Paoletti:2022gsn,Cruz:2023rmo,Seshadri:2000ky,Schleicher:2011jj,Pandey:2014vga, Bera:2020jsg,Bhaumik:2024efz, Bhaumik:2026mlc, Tashiro:2006uv, Mohapatra:2024djd, natwariya2020edges} for earlier work on this subject. In particular, our work improves on the recent analysis of~\cite{Cruz:2023rmo} in accounting for the effect of PMF energy losses on the PMF amplitude and in the statistical methods used to derive bounds from reionization. Here we perform a MCMC analysis using a neural network emulator, {\tt NNERO}~\cite{Facchinetti:2025hou}, that allows us to simulate the reionization history while varying both astrophysical parameters and PMF parameters inside the entire parameter space considered here. Another feature of our analysis is that we combine reionization constraints with galaxy UV luminosity function constraints, see also~\cite{Schleicher:2011jj}. 
 
Furthermore, we evaluate the sensitivity of the recently-built radio telescope Hydrogen Epoch of Reionization Array (HERA)~\cite{HERA:2021bsv,HERA:2021noe,HERA:2025ajm} to PMF heating and structure formation boost using a Fisher matrix analysis. Energy losses that affect the PMF energy density have already been shown to affect the PMF signature in earlier analyses~\cite{Bera:2020jsg, Bhaumik:2024efz}, which focused on nearly scale-invariant PMFs' imprint on the 21$\,$cm signal. Here we improve on those analyses with a more advanced treatment of the 21$\,$cm signal, focusing on HERA's sensitivity towards 21$\,$cm power spectrum, which we evaluate using {\tt exo21cmFAST}. Furthermore, we discuss degeneracies with other astrophysical parameters, and the dependence of the HERA sensitivity on the latter.

The rest of the paper is structured as follows. In Section~\ref{sec:prim-magn-fields}, we describe our modeling of PMFs, based on recent developments in the literature. We emphasize the role of PMFs in structure formation and their interplay with the IGM. In Section~\ref{sec:analys_and_results}, we detail our methodology to constrain PMFs from optical depth and UV luminosity function measurements. In Section~\ref{sec:21cm}, we discuss their imprint on the 21$\,$cm signal and forecast the expected sensitivity of HERA. Finally, we conclude in Section~\ref{sec:concl}. In the following, we keep $c$ and $\mu_0$ insertions in all analytical expressions to avoid confusion. The cosmological parameters are fixed to their best-fit values as reported by Planck (TT,TE,EE+lowE+lensing)~\cite{Planck:2018vyg}: $\Omega_{\rm cdm}h^2 = 0.1200$, $\Omega_{\rm b}h^2 = 0.02237$, $h = 0.6736$, $\ln(10^{10} A_{\rm s}) = 3.044$ and $n_{\rm s} = 0.9649$. The value of the optical depth to reionization will be discussed in Section~\ref{sec:opdep}.

\section{Primordial magnetic fields}
\label{sec:prim-magn-fields}
In this section, we introduce the PMF modeling and relevant notations used in this paper. We start by presenting some generalities before addressing their interplay with the IGM and then their impact on the matter power spectrum and the formation of structures.

\subsection{PMF power spectrum}

We express the time- and position-dependent magnetic field in terms of its value at recombination as:
 \begin{equation}
  \frac{ {\bf B}({\bf x},z)}{(1+z)^2 }= \chi_{\rm B}(z) \, \frac{{\bf B}({\bf x},z_{\rm rec})}{(1+z_{\rm rec})^2} \quad {\rm with}\quad \chi_{\rm B}(z_{\rm rec})=1\,.
   \label{eq:B(x,z)}
 \end{equation}
  $\chi_{\rm B}(z)$, a dimensionless measure of the PMF strength, captures the deviation from the na\"ive $(1+z)^2$ assumption.
 We introduce  the Fourier transform of the redshifted magnetic field at recombination  as
\begin{equation}
\frac{{\bf B}({\bf x},z_{\rm rec})}{(1+z_{\rm rec})^2} = \int
\frac{\dd^3k}{(2 \pi)^3}\,{{\bf B}}({\bf k}) \,e^{-i{\bf k}\cdot{\bf x}}\,,
\label{eq:Bk}
\end{equation}
where  ${{\bf B}}({\bf k})$ is the  Fourier dual to ${\bf B}({\bf x},z_{\rm rec})/(1+z_{\rm rec})^2$  and    ${\bf k}$ is the comoving wave vector.  The Fourier dual  can either be expressed in terms of its cartesian coordinates, $\{{\bf B}_i({\bf k})\}$ with $i=1,2,3$ or in terms of its projections onto the polarization unit vectors  $ \hat \epsilon_\alpha({\bf k})$ with $\alpha=1,2$, which are orthogonal to {\bf k} and to each other, with $ {{\bf B}}({\bf k})=\sum_\alpha B_\alpha ({\bf k})\hat \epsilon_\alpha({\bf k})$.
 The cartesian components of the Fourier dual of the magnetic field satisfy the two-point correlation function~\cite{Adi:2023doe}
 \begin{equation}
    \left\langle{\bf B}_i({\bf k}) {\bf B}_j^*\left({\bf q}\right)\right\rangle=\frac{(2 \pi)^3}{2} \delta^{(3)}\left({\bf k}-{\bf q}\right)\mathrm{P}_{ij}({\bf k}) P_{\rm B}(k)\label{eq:two-point-B}
\end{equation}
where $i,j=1,..,3$ denote the Cartesian coordinate indices, $\delta^{(3)}\left({\bf k}-{\bf q}\right)$ is a 3-dimensional Dirac delta and $\mathrm{P}_{ij}({\bf k})=\delta_{ij}-k_ik_j/k^2=\sum_\alpha \hat\epsilon_{\alpha i}({\bf k})\hat\epsilon_{\alpha j}^*({\bf k})$ is a projection
tensor resulting from the completeness relation of the polarization  vectors.

 We assume the redshift-independent PMF power spectrum of Eq.~(\ref{eq:two-point-B}) to scale as (see e.g.~\cite{Kim:1994zh})
 \begin{equation}
 P_{\rm B}(k)=A_{\rm B} k^{n_{\rm B}}\,,
 \label{eq:PB}
\end{equation}
where both $A_\B$ and $n_{\rm B}$ are constants.
According to Eq.~(\ref{eq:B(x,z)}), the redshift-dependent  PMF power spectrum -- associated to the variance of the dual to ${\bf B}({\bf x},z)/(1+z)^2$ -- is thus given by 
\begin{equation}
P_{\rm B}(k,z)=P_{\rm B}(k)   \chi_{\rm B}^2(z)\,.
    \label{eq:P_B(k,z)}
\end{equation}
Furthermore,  we will often quantify the PMF energy density by tracking the  magnetic power spectrum smoothed over a window of a certain scale   $\lambda_c=2\pi/k_c$. Here we use the  real-space Gaussian filtering and introduce the smoothed magnetic field strength $\sig_{\rm B}$ with
\begin{eqnarray}
    \sig_{\rm B}^2(\lambda_c,z) &=& \int \frac{ \dd k^3}{(2\pi)^3} \, P_{\rm B}(k,z)\, e^{-2k^2/k_c^2}\cr
    &=& \chi_{\rm B}(z)^2 \frac{A_{\rm B} }{(2\pi)^2 }\left(\frac{k_c}{\sqrt2}\right)^{3+n_{\rm B}}\Gamma\left[\frac{3+n_{\rm B}}{2}\right]\,.
\label{eq:sigma_B}
\end{eqnarray}
For further convenience, we parametrize the PMF energy density amplitude  with $s_{\rm B}$, defined as the magnetic field strength in the absence of PMF decay redshifted from recombination until today, smoothed over a Gaussian filter of  $\lambda_c= 1 \,{\rm Mpc}$,
\begin{equation}
\sig_{\rm B}(1\, {\rm Mpc}, z)\equiv {s_{\rm B}} \chi_{\rm B}(z)  \quad {\rm with} \quad
    s_{\rm B} \equiv \sig_{\rm B}(1\,{\rm Mpc}, z_{\rm rec}) \, .
    \label{eq:chiBsB}
\end{equation}
In the rest of this paper, the PMF parameters are thus 
\begin{equation}
    \{s_{\rm B}, n_{\rm B}\}
    \label{eq:PMFparams}
\end{equation}
to characterize the normalization and the tilt of the PMF power spectrum. In the previous literature, the magnetic field amplitude has been characterized by $\sig_\B(1\,{\rm Mpc},z_0)$ often denoted with $\sigma_{{\rm B},0}$, and mostly assumed to be equal to $s_{\rm B}$. With  the treatment of the magnetic field energy losses considered here, we have 
\begin{equation}
    \sigma_{{\rm B},0}\equiv\sig_{\rm B}(1\,{\rm Mpc},0)=s_{\rm B}\chi_{\rm B}(0)\neq s_{\rm B} 
    \label{eq:sigmaB0}
\end{equation}
 the deviation from the $s_{\rm B}$ normalization being captured by the $\chi_{\rm B}(0)$ factor that depends on the IGM evolution history as discussed in the next sections.
%

\subsection{IGM evolution equations }
\label{sec:IGMevol}

In the early universe, before recombination, the photons exert a drag force on the cosmic plasma that affects both the evolution of cosmic plasma and  the PMF, see~\cite{Jedamzik:1996wp, Subramanian:1997gi, Subramanian:2015lua}.  Beyond recombination, this results in the damping of  PMFs with wave modes larger than the comoving Alfv\'{e}n wave number, $k_{\rm A}$, with  
\begin{equation}
    k_{\rm A} \equiv k_\gamma(z_{\rm rec})/ v_{\rm A}(z_{\rm rec})\quad {\rm and} \quad v_{\rm A}(z)=  \frac{\sig_{\rm B}(\lambda_{\rm A}, z)} {\sqrt{\frac{4}{3}\mu_0 \rho_\gamma(z_0)}}\,,
    \label{eq:kAvA}
    \end{equation}
    where $k_\gamma(z_{\rm rec})$ denotes the photon diffusion scale at recombination while the Alfv\'{e}n velocity squared, $v_{\rm A}^2$, scales as the  ratio between the smoothed magnetic field strength on a scale $\lambda_{\rm A}=2\pi/k_{\rm A}$ and the photon pressure plus energy density contribution today, see e.g.~\cite{Kunze:2013uja}. We obtain 
  \begin{equation}
    k_{\rm A} = \frac{271 \sqrt{2} \, {\rm nG} }{\sig_{\rm B}(\lambda_{\rm A},z_{\rm rec})} \,{\rm Mpc}^{-1}
    \label{eq:kAnum}
    \end{equation}
    using Planck 2018 best-fit cosmological parameter values. This is
    in good agreement with the values reported in~\cite{Kunze:2013uja} (the  factor $\sqrt{2}$ is due to our choice of normalization of $\sig_{\rm B}$).
    As a result, after recombination, the scale-dependent magnetic field power
spectrum is exponentially suppressed at large wavenumbers $k$ and reduces to
\begin{equation}
 P_{\rm B}(k)= A_{\rm B} k^{n_{\rm B}} e^{-2k^2/k_{\rm A}^2}\,.
 \label{eq:PBAlfven}
\end{equation}
In what follows, all insertions of the   magnetic power spectrum come with an $e^{-2k^2/k_{\rm A}^2}$ suppression factor as the physics involved is post-recombination.

After recombination, PMFs can affect the IGM evolution in multiple ways, see e.g.~\cite{Subramanian:2015lua} for a review.
On the one hand, {\it ambipolar diffusion} (AD) refers to the magnetic field-induced drift of the charged components of the plasma  while leaving the neutral atoms unaffected. This creates a relative drift between the charged and neutral particles, which induces friction that converts magnetic field energy into heat within the IGM. On the other hand, as recombination increases the photons' mean free path, the fluid’s viscosity decreases, enabling the development of {\it decaying turbulences} (DT) in the plasma.
The evolution of the IGM (or gas) temperature, $T_{\tt k}$, the ionized fraction of hydrogen, $x_\e$, and the PMF energy density, characterized here with $\chi_{\rm B}$, is determined by a set of coupled equations~\cite{Jedamzik:1996wp, Sethi:2004pe, Banerjee:2004df, Pinto:2008zn}
\begin{eqnarray}
  \frac{\partial T_{\rm k}}{\partial z} &=& -\frac{2}{3k_{\rm B}n_{\rm b}}\frac{1}{(1+z)H}\sum_{\beta}\epsilon^\beta_{\rm heat} + \frac{2}{3}\frac{T_{\rm k}}{n_{\rm b}} \frac{{\rm d} n_{\rm b}}{{\rm d} z} - \frac{T_{\rm k}}{1+x_\e}\frac{{\rm d} x_\e}{{\rm d}z}\label{eq:Tkdot}\\
  \frac{\partial x_\e}{\partial z}  &=& -\frac{1}{(1+z)H}\left[\Lambda_{\rm ion} - \Lambda_{\rm rec}\right] \label{eq:xedot}\\
     \frac{{\rm d} \chi_{\rm B}}{{\rm d}z} &= &\frac{1}{2\chi_{\rm B}} \frac{1}{(1+z) H} \left[\overline{\gamma_{\rm AD}} +\overline{\gamma_{\rm DT}} \right]  \, ,
     \label{eq:chiBdot}
\end{eqnarray} 
see also e.g.~\cite{Chluba:2015lpa, Paoletti:2018uic, Cruz:2023rmo,  Minoda:2022nso,Bhaumik:2024efz}. In these equations, $k_{\rm B}$ is the Boltzmann constant, $ H=H(z)$ is the Hubble rate,  $\Lambda_{\rm ion}$  and $\Lambda_{\rm rec}$ denote the ionization and recombination rates,
and
$\epsilon^\beta_{\rm heat}$ is the heating rate from each source $\beta$, including Compton scattering (effective at high redshifts, $z \gtrsim 300$) and heating from X-rays ($\epsilon_{\rm heat}^{\rm X}$).  
Losses of magnetic field energy  density due to ambipolar diffusion and decaying turbulences constitute additional sources of  heating, that we denote with  $\epsilon_{\rm heat}^{\rm AD}$ and  $\epsilon_{\rm heat}^{\rm DT}$. 
The $\gamma_{\rm AD}$ and $\gamma_{\rm DT}$ energy loss rates entering in the magnetic field evolution, are related to the IGM heating rates of Eq.~(\ref{eq:Tkdot}) as follows:
\begin{equation}
    \gamma_{\rm AD/DT} \equiv \frac{2\mu_0}{\sig_{\rm B}^2(\lambda_A, z_{\rm rec})}\epsilon^{\rm AD/DT}_{\rm heat}(1+z)^{-4},
    \label{eq:gammaADDT}
\end{equation}
where the overlines in Eq.~(\ref{eq:chiBdot})  indicate spatial averages.

The rate of AD energy injection is driven by the magnetic field-induced Lorentz friction force
and is estimated as~\cite{Sethi:2004pe,  Pinto:2008zn,   Kunze:2013uja,  Chluba:2015lpa, Paoletti:2018uic}: 
\begin{eqnarray}
    \epsilon_{\rm heat}^{\rm AD}=\frac{c^2}{\mu_0^2} \frac{1-x_\e}{x_\e}  \frac{2 m_{\rm H}}{\rho_{\rm b}^2 \left<\sigma v\right>_{\rm HH^+}}(1+z)^2\langle|{\bf B}\times (\nabla\times {\bf B})|^2 \rangle 
    \label{eq:epsheatAD}
\end{eqnarray}
where $\langle ..\rangle$ denotes the ensemble average, $\nabla$ is  is the divergence with respect to the comoving position,  $m_{\rm H}$ is the hydrogen mass, and $\left< \sigma v\right>_{\rm HH^+} \simeq 6.5 \times 10^{-10}\,(T_{\rm k}/{\rm K})^{0.375} \, {\rm cm^3}/{\rm s}$ is the velocity-averaged cross-section between neutral and ionized hydrogen atoms as a function of the IGM temperature~\cite{Pinto:2008zn, Schleicher:2008aa}. Just with the magnetic field redshift dependence in the numerator, we expect already a $\epsilon_{\rm heat}^{\rm AD}\propto\chi_{\rm B}(z)^4$ dependence.
In agreement with the recent recalculation of the Lorentz force average over the distribution of~\cite{Adi:2023doe}, we have checked that 
\begin{eqnarray}
     \frac{\langle|{\bf B}\times (\nabla\times {\bf B})|^2 \rangle}{(1+z)^8} &=&\frac14 \int \frac{\dd^3k}{(2\pi)^3} \frac{\dd^3k_1}{(2\pi)^3} P_{\rm B}(k_1,z) P_{\rm B}(k_2,z) \left(k^2 + (k^2 - 2k k_1 \mu)\mu^2\right)\cr
     &= &k_{\rm A}^2\sig_{\rm B}^4(\lambda_{\rm A}, z_{\rm rec})  \chi_{\rm B}^4\times f_{\rm L}(n_{\rm B}+3) \, \label{eq:Lforcesq}\\
     \hspace{-1cm}&{\rm with }&  {\bf \hat k}_1\cdot {\bf \hat k}= \mu\;,  {\bf k}={\bf k}_1+{\bf k}_2\;{\rm and}  \;f_{\rm L}(x) \simeq  \frac{4.2  x^{1.92}}{1+ 40 x^{0.85}}\, \label{eq:fL} 
\end{eqnarray}
where ${\bf k,  k}_i$ are the wavenumber 3-vectors while ${\bf \hat k, \hat k}_i$ denote the corresponding unit vectors.   The $n_{\rm B}$ dependence can be fitted with a simple function $f_{\rm L}(n_{\rm B}+3)$ (see also~\cite{Chluba:2015lpa, Paoletti:2018uic}).

Furthermore, MHD simulations in flat space have been used to estimate the decay rate of PMFs in a matter-dominated universe as~\cite{Banerjee:2004df, Sethi:2004pe}
\begin{eqnarray}
    \overline{\gamma_{\rm DT}} &=& - m \frac{H(z)}{\ln [rq(z)]} \chi_{\rm B}^2 \label{eq:gammaDT}
\end{eqnarray}
where we have inserted the $\chi_{\rm B}^2$ factor to account for magnetic field decays, $m \equiv 2(n_{\rm B}+3)/(n_{\rm B} + 5)$, $q(z) \equiv (1+z)/(1+z_{\rm rec})$, and 
  \begin{equation}
     r \equiv \left(1+\frac{t_{\rm d}}{t_{\rm rec}}\right)^{-2/3} \quad {\rm with} \quad   \frac{t_{\rm d}}{t_{\rm rec}} 
     = \left(\frac{k_{\rm J}}{k_A}\right)^{\frac{5+n_{\rm B}}{2}} \, .
\end{equation} 
The comoving magnetic Jeans wave number,  $k_{\rm J}$,  beyond which magnetic pressure gradients counteract the
gravitational collapse,  enters in the above equation and is evaluated as~\cite{Sethi:2004pe}
\begin{equation}
    k_{\rm J}^2=\frac{4\pi G_{\rm N} \mu_0 \overline\rho_{{\rm m}, 0}\overline \rho_{{\rm b}, 0}}{\gamma^2 \Sigma_{\rm B}^2(\lambda_{\rm J},z_{\rm rec})}\,,
    \label{eq:kJ}
\end{equation}

where $\overline\rho_{\rm m,0}$ and $\overline\rho_{\rm b,0}$ are the energy densities today of matter and baryons respectively and $\lambda_{\rm J}=2\pi/k_{\rm J}$. We take  $\gamma=4/5$  as in~\cite{Cruz:2023rmo} based on simulations, see~\cite{Kim:1994zh}. Equivalently, using Eqs.~(\ref{eq:sigma_B}) and~(\ref{eq:chiBsB}), we have:
\begin{equation}
    \frac{k_{\rm J} }{2\pi\,{\rm Mpc}^{-1}}=\left(\frac{\sqrt{2 G_{\rm N}\, \mu_0 \overline\rho_{{\rm m}, 0}\overline\rho_{{\rm b}, 0}}}{\gamma\, s_{\rm B}\times \sqrt{2 \pi}\,{\rm Mpc}^{-1}} \right)^{\frac{2}{5+n_{\rm B}}}\simeq \left( \frac{2.1\, {\rm nG}}{s_{\rm B} }  \right)^{\frac{2}{5+n_{\rm B}}}\,. 
     \label{eq:kJnum}
\end{equation}
Note that the expression of Eq.~(\ref{eq:gammaDT}) (from~\cite{Sethi:2004pe}) differs from the one  sometimes found in literature (see e.g. \cite{Chluba:2015lpa, Kunze:2013uja}), where the decay of the magnetic field strength is incorporated directly into the formula. In our case, this effect is instead captured by the definition of $\chi_{\rm B}$ and its evolution equation, Eq.~(\ref{eq:chiBdot}).

In our treatment of the high-redshift evolution of the IGM with {\tt Hyrec}, $x_\e$ denotes the total ionized fraction and both $\Lambda_{\rm ion}$ and $\Lambda_{\rm rec}$   are functions of the temperatures, and thus of the redshift, see~\cite{Ali-Haimoud:2010hou, Chluba:2015lpa}. For low redshifts, we use {\tt exo21cmFAST} that treats separately the ionized fraction in mostly neutral IGM and in mostly ionized regions~\cite{Mesinger:2010ne}.  The former corresponds to $x_\e$  set by Eqs.~(\ref{eq:Tkdot})-(\ref{eq:chiBdot}),  evolved as a function of position and time, as is the case of $T_{\rm k}$ in {\tt exo21cmFAST}. For the ionized fraction in mostly ionized regions at late times,  the inhomogeneous reionization history  is obtained through an excursion set algorithm that evaluates the volume filling factor of HII regions as $Q_{\rm HII}(z)$.  Considering ionization in both the mostly-ionized and mostly-neutral IGM, we define
the density-weighted averaged ionized fraction as:
\begin{equation}
    \overline{x_i}\equiv \overline{x_{\rm HII}(1+\delta_{\rm b})}\,, 
    \label{eq:xHIIdb}
\end{equation}
where the total ionized fraction $x_{\rm HII}$ is obtained in {\tt exo21cmFAST} combining both $x_\e$ and $Q_{\rm HII}$ contributions ($ \overline{x_{\rm HII}}\approx Q_{\rm HII} + \overline{x_\e} (1-Q_{\rm HII})$) and
$\delta_{\rm b}$ denotes the baryon overdensities. 
Equation~(\ref{eq:xHIIdb}) accounts for spatial correlations between $x_{\rm HII}$  and $\delta_{\rm b}$. In particular, $\overline{x_i}\neq  \overline{x_{\rm HII}}\overline{(1+\delta_{\rm b})}$, as emphasized in~\cite{Liu:2015txa} (see also~\cite{Shmueli:2023box,Facchinetti:2025hou}). \repl{In this work, we set the transition redshift between {\tt Hyrec} and {\tt exo21cmFAST} to $z=35$. This is the default redshift at which  {\tt 21cmFAST} simulations are initialized. We have explicitly checked that  increasing the initialization redshift negligibly changed our results.~\footnote{\repl{Specifically, we ran lightcones at $(n_{\rm B},s_{\rm B}/{\rm nG})=\{(-2.5, 0.3), (-2.0, 0.15), (-1.5, 0.04),(-1.0, 0.01)\}$, all chosen along the Planck-$\tau$ exclusion contour of Fig.~\ref{fig:constraints}, with both $z_{\rm transition}=35$ and $z_{\rm transition}=45$ and found that the resulting optical depths change by $0.28\%, 0.55\%, 1.8\%$ and $0.41\%$ respectively.  This  negligibly impacts our derived constraints.}}}

PMFs affect $x_{\rm HII}$ in two ways. First, through energy injection, see Eqs.~(\ref{eq:Tkdot})--(\ref{eq:chiBdot}). \repl{ Note that in this work we only consider energy injection in mainly neutral regions,  where $x_e$ and $T_k$ are computed. }\footnote{ \repl{ Explicitly computing the energy injection in mainly ionized regions could further increase the total ionized fraction. This   would however necessitate dedicates MHD simulations which is beyond the scope of this work.}} Second, as  we will see in the next section, they  enhance the growth of perturbations at small scales, providing an additional modification of the IGM evolution history that affects the ionized fraction in mainly ionized regions.

\subsection{Boost of small scale structures}
\label{sec:growth}

Primordial magnetic fields induce density perturbations, which grow due to gravitational instabilities and cause early structure formation in the universe at small scales~\cite{Subramanian:1997gi, Kim:1994zh,  Jedamzik:1996wp}. For the  PMF parameter space of relevance here, this enhanced growth  of matter overdensities typically affects   a range of halo masses  $\gtrsim 10^6 {\rm M}_\odot$, see Sec.~\ref{sec:HMF} below.\footnote{More recently,~\cite{Ralegankar:2023pyx} showed that PMFs can also induce the formation of mini-halos with mass $\ll 10^6 {\rm M}_\odot$   (see also~\cite{Olea-Romacho:2025qag} for interesting applications to dark matter). In this paper, we neglect those extra contributions focusing on a single population of galaxies. } It can be shown that after recombination, when the photon density drops and the baryons' coupling to photons get suppressed, the Lorentz force induced by the PMFs acts as an extra source term in the Euler equation for the baryon velocity, ${\bf v}_{\rm b}$,  
\begin{equation}
 \dot{\bf v}_{\rm b} +H {\bf v}_{\rm b}=- (1+z){\bf\nabla}\psi-  {\bf S}_L\quad {\rm with}\quad  {\bf S}_L= (1+z)\frac{{\bf B}\times(\nabla\times{\bf B})}{\mu_0 \rho_{\rm b}}\,,
    \label{eq:sourceEulerb}
\end{equation}
where $\psi$ is the gravitational potential and the dot denotes the derivative w.r.t. the proper time $t$. The Lorentz force therefore sources additional matter density perturbations.

\subsubsection{Growth of linear matter perturbations}
\label{sec:lingrowth}

The equation
governing the resulting growth of matter perturbations at linear order takes the form~\cite{1978ApJ...224..337W, Kim:1994zh, 2012PhRvD..86d3510S, 2012JCAP...11..055F, Pandey:2012ss, Adi:2023qdf, Ralegankar:2024arh}
\begin{equation}
    \ddot\delta_{\rm m} + 2H\dot \delta_{\rm m} - 4\pi G_{\rm N}\overline\rho_{\rm m}\delta_{\rm m} =  (1+z)f_{\rm b} \nabla\cdot{\bf S}_L\, ,
    \label{eq:growtheq}
\end{equation}
where $\delta_{\rm m}=\delta_{\rm dm} f_{\rm dm}+\delta_{\rm b}f_{\rm b}$ denotes the total matter density perturbations (DM + baryons) and $f_i=\overline\rho_{i,0}/\overline\rho_{\rm m,0}$ give the relative DM and baryon contribution with respect to the total matter density for $i=$ dm and b respectively,  and $G_{\rm N}$ is the Newton constant. The extra PMF contribution on the right hand side induces an enhancement of the matter power spectrum on scales between the magnetic Jeans scale  and the Alfv\'en scale.

The homogeneous solution to the matter perturbation growth equation of Eq.~(\ref{eq:growtheq}) is the $\Lambda$CDM linear matter density evolution, $\delta_{\rm m}^{\Lambda \rm  CDM } ({\bf{x}},z)$. We denote the special solution to the full equation with $\delta_{\rm m}^{\rm B} ({\bf{x}},z)$ so that the total matter perturbation can be obtained as  $\delta_{\rm m} ({\bf{x}},z)= \delta_{\rm m}^{\Lambda \rm  CDM } ({\bf{x}},z)+\delta_{\rm m}^{\rm B} ({\bf{x}},z)$. Furthermore, $\delta_{\rm m}^{\rm B}$ is decomposed as
\begin{eqnarray}
    \delta^{\rm B}_{\rm m} ({\bf x},t)&=& M_{\rm B}(t)\Delta({\bf x})  \label{eq:deltas} \\
    {\rm with}\;\; \Delta({\bf x})&=&\frac{f_{\rm b}}{\mu_0 \overline\rho_{{\rm b},0}} \frac{\nabla\cdot({\bf B}({\bf{x}},z_{\rm rec})\times(\nabla\times{\bf B}({\bf x},z_{\rm rec})))}{ (1+z_{\rm rec})^4}\,,
    \label{eq:Deltax}
\end{eqnarray}
such that the magnetic growth function
$M_{\rm B}(t)$, satisfies:
\begin{equation}
    \ddot M_{\rm B} + 2H\dot M_{\rm B} - 4\pi G_{\rm N} M_{\rm B} = (1+z)^{3} \chi_{\rm B}^2\quad {\rm with}\quad\dot M_{\rm B}(t_{\rm rec})=M_{\rm B}(t_{\rm rec})=0\,,
    \label{eq:MBgrowth}
\end{equation}
where $\dot M_{\rm B}(t)= \dd M_{\rm B}(t)/\dd t$  denotes the  proper time derivative. 
It can be shown that the power spectrum of the dual to  $\Delta({\bf x})$ is given by~\cite{Adi:2023doe}:
\begin{eqnarray}
 P_{\Delta}(k)&=&\left(\frac{f_{\rm b}k }{4\pi\mu_0 \overline\rho_{{\rm b},0}}  \right)^2\int_0^\infty k_1^2\dd k_1\int_{-1}^1 \dd\mu \,P_{\rm B}(k_1) P_{\rm B}(k_2) \left(k^2 + (k^2 - 2k k_1 \mu)\mu^2\right)\,, \label{eq:PDelta}
\end{eqnarray}
where we use the same relations between $\mu, k_1, k_2$ and $k$ as in the computation of the ensemble average of the squared  Lorentz force of Eq.~(\ref{eq:fL}).  In this case, $P_{\Delta}(k)$ is expected to scale as $k_A \Sigma_{B}^4(\lambda_A,z_{\rm rec})$ when setting $k
=k_A$.
The power spectrum of the total matter perturbations  takes the form
\begin{equation}
P_{\rm m}(k, t) = D^2(t) P_{\rm m}^{\Lambda\rm CDM} (k) + M^2_{\rm B}(t)\frac{P_\Delta(k)}{(1+(k/k_{\rm J})^2)^2}\,,
\label{eq:matterPS}
\end{equation}
assuming no correlation between the $\Lambda$CDM
and the magnetically induced matter density contrasts as in e.g.~\cite{Adi:2023doe, Adi:2023qdf,Cruz:2023rmo, Ralegankar:2024arh}.  $D(t)$ and $P_{\rm m}^{\Lambda\rm CDM} (k)$ denote the $\Lambda$CDM growth function and matter power spectrum while the $(1+(k/k_{\rm J})^2)^{-2}$ factor accounts for magnetic Jeans damping, see~\cite{Cruz:2023rmo}. The difference with respect to the treatment in previous works~\cite{Adi:2023doe, Adi:2023qdf, Cruz:2023rmo} is the insertion of $\chi_{\rm B}$ appearing as a weighting factor in the $(1+z)^{3}$ source term of Eq.~(\ref{eq:MBgrowth}). This decreases the enhancement of the matter power spectrum at small scales w.r.t. previous analyses.

\begin{figure}[t]
    \centering
    \includegraphics[width=0.99\linewidth]{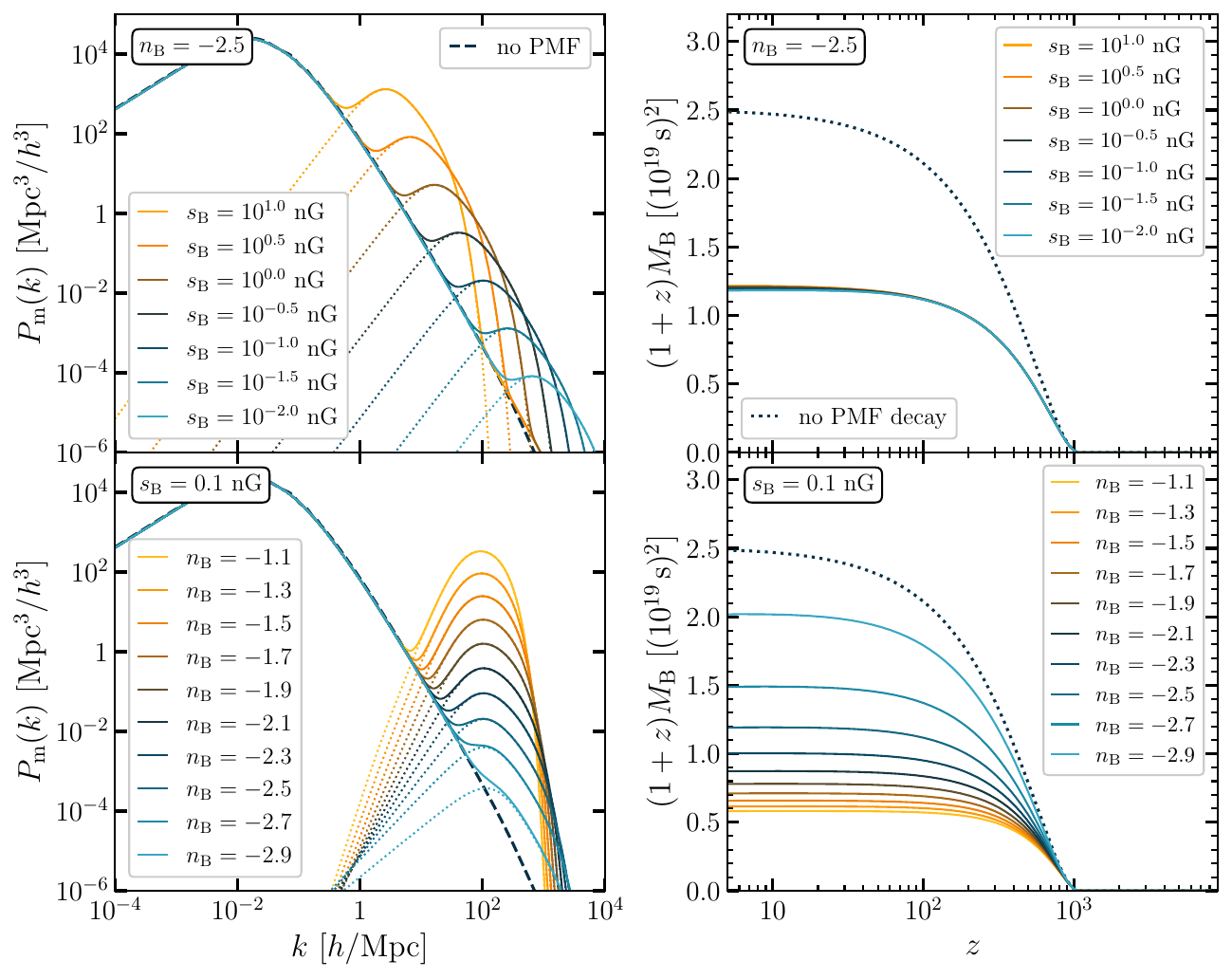}
    \caption{{\bf Left panels.} Matter power spectrum as a function of wavenumber for different values of the PMF normalization $s_{\rm B}$ (top panels) and tilt parameter $n_{\rm B}$ (bottom panels).
     Solid lines show the total power spectrum while dotted line show the PMF contribution alone. The black dashed line show the $\Lambda$CDM power spectrum (that is, without PMF). {\bf Right panels.} Corresponding evolution of the PMF growth function.  The black dotted line shows the magnetic growth function neglecting the PMF energy losses.}
    \label{fig:ps_pmf}
\end{figure}

In Fig.~\ref{fig:ps_pmf}, we show the resulting  matter power spectrum (left column) and the corresponding magnetic  growth function (right column) for different values of the PMF normalization $s_{\rm B}$ (top panels) and tilt parameter $n_{\rm B}$ (bottom panels). Our results are qualitatively similar to those of~\cite{Cruz:2023rmo}, displaying an increased amplitude of the enhancement of the small scale power spectrum for larger normalization and tilt parameters. The Jeans and Alv\'en damping also enter at lower $k$ values when increasing $s_{\rm B}$ or $n_{\rm B}$ as expected from Eqs.~(\ref{eq:kAnum})  and~(\ref{eq:kJnum}). The suppression of the PMF enhancement due to PMF energy losses can be extracted from the right plots by comparing the dotted line, which represents the magnetic growth function when neglecting those losses, with the colored continuous lines that we obtain from the full treatment for fixed $n_{\rm B}$ and $s_{\rm B}$ in the top and bottom panels, respectively. This suppression appears to be almost independent of $s_{\rm B}$ at fixed $n_\B$ while increasing values of $n_{\rm B}$ at fixed $s_\B$ induce stronger suppression. As already pointed out in~\cite{Chluba:2015lpa}, DT dominate energy losses at early times in Eq.~(\ref{eq:chiBdot}) and dominate the source of energy losses for the scenarios illustrated here. The resulting PMF strength suppression thus mainly depends on $n_{\rm B}$. When $n_{\rm B}\to -3$ however, DT losses are suppressed, which explains why the line for $n_{\rm B}=-2.9$ (light blue) is closest to the dotted line.  Also note that the results presented here differ from~\cite{Cruz:2023rmo} in the case of no energy losses. We suspect the origin of some difference to be labeling typos in their Fig.~2 (otherwise inconsistent). We show a comparison in Appendix~\ref{app:PSboost}.

\subsubsection{Enhanced halo number density}
\label{sec:HMF}

\begin{figure}
    \centering
    \includegraphics[width=0.75\linewidth]{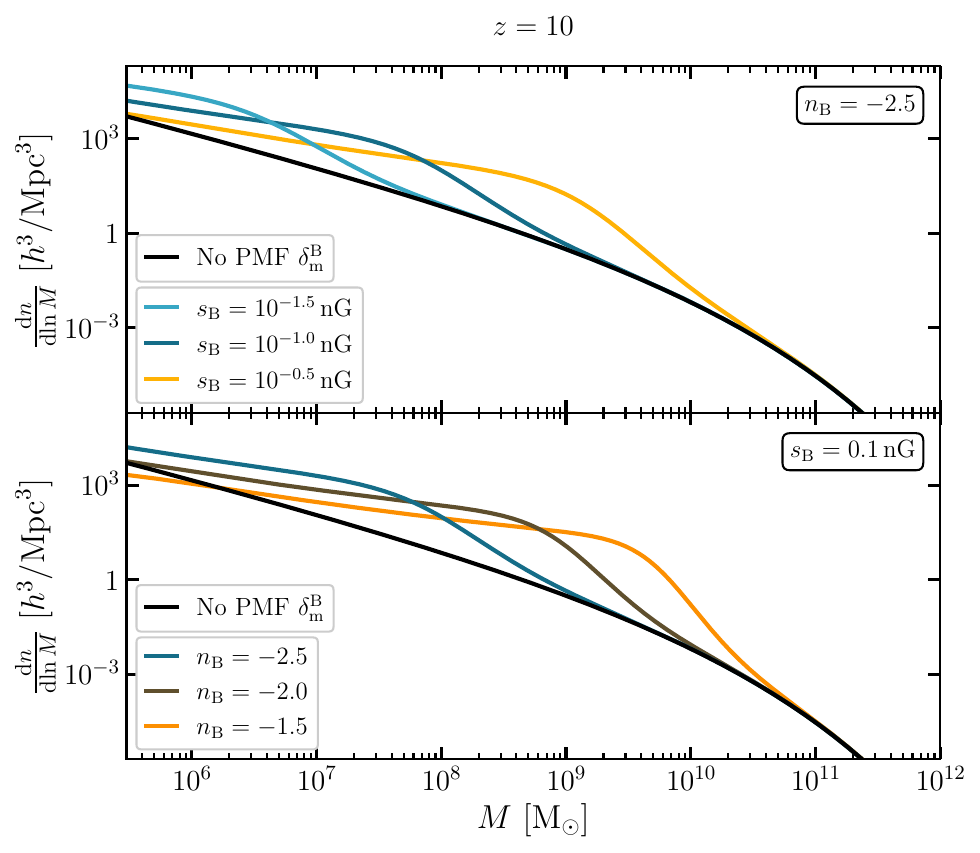}
    \caption{\textbf{Upper panel:} The halo mass function for the CDM prediction (black) and different values of $s_{\rm B}$ for a fixed $n_{\rm B}=-2.5$. \textbf{Lower panel:} same, but for different values of $n_{\rm B}$ and a fixed value of $s_{\rm B}=0.1\,{\rm nG}$.}
    \label{fig:hmf_nb_sb}
\end{figure}
The enhancement of power at large wavenumbers affects the formation of structures. In this work, we include this effect through its impact on the halo mass function (HMF), ${\rm d}n/{\rm d}M$, which counts the halo number density per unit mass. In {\tt exo21cmFAST}, it is evaluated from the linear matter power spectrum using the excursion set formalism, which leads to an analytical expression for the HMF by studying the Brownian motion of the linear density contrast, see e.g. Press-Schechter \cite{Press:1973iz} and Sheth-Tormen \cite{Sheth:1999mn}. To that end, we compute the Sheth and Tormen HMF incorporating the variance of the total linear matter power spectrum of Eq.~(\ref{eq:matterPS}) within a spherical region in Fourier space using a sharp-$k$ window function. We use the sharp-$k$ window instead of the default top-hat window function, as it was shown to properly account for small-scale modifications to the matter power spectrum (as e.g. in the presence of warm dark matter~\cite{Benson:2012su, Schneider:2013ria}). All details on our halo mass computation can be found in~\cite{Dandoy:2025twl}. In Fig.~\ref{fig:hmf_nb_sb}, we illustrate the effect of PMFs on the HMF at $z=10$ for fixed $n_{\rm B}=-2.5$ (top) or $s_{\rm B}=0.1\,$nG (bottom). The HMF is enhanced in a fixed range of halo masses corresponding to mass scales approximately set by the Jeans and Alfv\'en scales. For larger normalization and tilt parameters, the enhancement is stronger and appears at higher halo masses (that correspond to smaller wavenumbers $k$).

The increased halo abundance due to PMFs can be expected to affect the reionization history, the UV luminosity function and the $21\,$cm signal, as already discussed in e.g.~\cite{Schleicher:2011jj, Cruz:2023rmo, Katz:2021iou, Bhaumik:2026mlc}. To evaluate the PMF imprint, we use {\tt exo21cmFAST}, assuming one single population of galaxies during Cosmic Dawn (CD) and following the prescriptions introduced in~\cite{Mesinger:2010ne,Park:2018ljd}. The number of ionizing photons per baryon depends on the HMF as 
\begin{equation}
    n_{\rm ion} = \frac{N_{\gamma / {\rm b}}}{\overline\rho_{\rm b}(z)}\int_0^\infty {\rm d}M\,\frac{{\rm d}n}{{\rm d}M} f_{\rm duty} M_\star  f_{\rm esc},
\end{equation}
where $N_{\gamma / {\rm b}}$ is the number of ionizing photons per stellar baryon fixed to 5000, $f_{\rm duty}(M)= \exp(-M_{\rm turn}/M)$ is the duty cycle function, which accounts for the threshold mass for star-forming galaxies, $M_{\rm turn}$, and $f_{\rm esc}$ is the fraction of ionizing photons that escape galaxies (see Sec.~\ref{sec:analys_and_results} for more details). $M_\star(M)$ is the stellar mass of a galaxy in a host halo of mass $M$. This is of particular relevance when estimating the constraints from the optical depth to reionization in Sec.~\ref{sec:an-reio}. \repl{ Indeed, the computation of  the volume filling factor
of HII regions, $Q_{\rm HII}$, directly depends on $n_{\rm ion}$ and thus of the HMF.  $Q_{\rm HII}$  is computed by  
{\tt exo21cmFAST} with an  excursion-set algorithm~\cite{Furlanetto:2004nh} to compute inhomogeneous reionization. Specifically, a cell is considered ionized when
$n_{\rm ion} \geq (1 + n_{\rm rec})(1-x_e)$
where
$n_{\rm rec}$ gives the cumulative number of recombinations per baryon within spheres of decreasing
radii and $x_e$ accounts for  ionization in mostly-neutral IGM. }
Furthermore, the specific X-ray emissivity at position ${\bf x}$, redshift $z$ and with energy $E$, which substantially shapes the form of the $21\,$cm signal during the epoch of heating, also depends on the HMF. It is implemented as
\begin{equation}
    \epsilon_\X({\bf x}, E, z) = [1+\overline \delta_{\rm b}({\bf x}, z)] \frac{ \mathcal{L}_\X E^{-\alpha_\X}}{t_\star H^{-1}(z)}\int_0^\infty  {\rm d}M\,\frac{{\rm d}n}{{\rm d}M} f_{\rm duty} M_\star  \,,
    \label{eq:epsX}
\end{equation}
where $t_\star$ is a free parameter between zero and one that sets the characteristic time scale of star formation with respect to the Hubble time, $\overline\delta_{\rm b}({\bf x}, z)$ is the average value of the nonlinear matter density contrast of the shell around $({\bf x}, z)$, and $\mathcal{L}_\X^i$ is the normalization of the X-ray luminosity per star formation rate (SFR).  The exponent $\alpha_{\rm X}$ remains fixed to its fiducial value, $\alpha_{\rm X}=1.0$.  The latter is conventionally expressed in terms of  the integrated soft-band ($E_0 < E < 2~{\rm keV}$) luminosity per SFR (in units of $\rm erg \,yr\,s^{-1}\,{\rm M}_\odot^{-1}$) 
\begin{equation}
   L_\X =  \mathcal{L}_\X \int_{E_0}^{2\,{\rm keV}} {\rm d } E \,E^{-\alpha_\X}\,,
   \label{eq:LX}
\end{equation}
where $E_0$  is the threshold energy for X-rays to escape the host galaxy (and thus heat the IGM). The X-ray heating contribution evaluated from Eq.~(\ref{eq:epsX}) is particularly relevant to estimate the sensitivity of $21\,$cm cosmology experiments, see Sec.~\ref{sec:21cm}.

\begin{figure}
    \centering
    \includegraphics[width=0.5\linewidth]{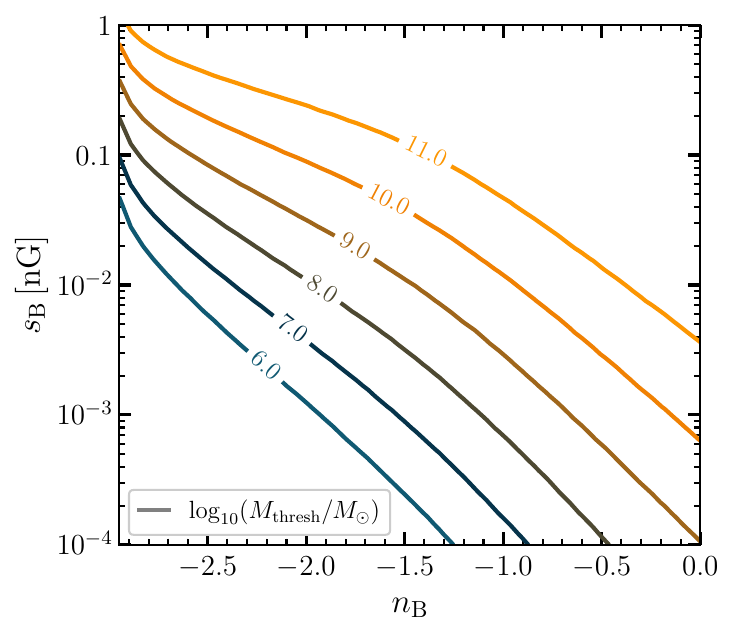}
    \caption{Values of $\log_{10}(M_{\rm thresh}/{\rm M}_\odot)$ as a function of $n_{\rm B}$ and $s_{\rm B}$. $M_{\rm thresh}$ indicates the largest halo mass for which the PMF-induced enhancement of the HMF equals $20\%$. }
    \label{fig:hmf_mthresh_contour}
\end{figure}

Figure~\ref{fig:hmf_mthresh_contour} shows the value of $M_{\rm thresh}$ in the $(n_{\rm B},s_{\rm B})$ plane. We define $M_{\rm thresh}$ as the largest halo mass at which the PMF-induced enhancement of the halo mass function is at least 20\%, in other words $\left.\frac{\dd n}{\dd M}\right._{\rm }/\left(\frac{\dd n}{\dd M}\right)_{\rm no~PMF~\delta_m^B}\geq1.2$. This mass is essentially set by the comoving magnetic Jeans scale $k_{\rm J}$. In our simulations, the HMF always appears multiplied by the duty cycle, which leads to a suppression of the emission of ionizing, UV and X-ray radiation at halo masses below $M_{\rm turn}$. If the threshold mass for star-forming galaxies is much larger than the threshold mass for PMF enhancement, $M_{\rm turn}> M_{\rm thresh }$, the increased density of low-mass halos as induced by PMFs is not expected to leave a measurable imprint on the observables considered here.

More details on the definitions of the astrophysical quantities introduced here can be found in~\cite{Park:2018ljd}. The range of astrophysical parameters considered in the UV LF and reionization analysis are provided in the fourth column of Tab.~\ref{tab:astro_params}. Our plots are based on the fiducial parameters given in the last column of the table. This parameter combination was shown to yield an optical depth to reionization $\tau$ consistent with the best-fit to Planck 2018 data in~\cite{Qin:2020xrg}.

Previously,~\cite{Cruz:2023rmo} also took into account the enhancement at small scales in $P_{\rm m}$, yet we understand that they use the default window function in {\tt 21cmFAST} (a top-hat function in position space) and consider two populations of galaxies, which primarily host PopII and PopIII stars respectively. Here we consider one single population of galaxies and a sharp-$k$ cut-off window function. Importantly, the former choice allows us to correct for the photon non-conservation inherent to the excursion-set formalism employed by such semi-analytical codes \cite{Park:2021eux} (i.e. to run our simulations with the \texttt{PHOTON\_CONS} option set to \texttt{True}).\footnote{To our knowledge, in {\tt 21cmFAST} the photon conservation algorithm is only implemented in the case of a single population of galaxies. } This allows us to compute the optical depth to reionization meaningfully.

\repl{It should be emphasized that the methodology used here to evaluate the impact of PMF on non-linearly evolved matter perturbations stems from linear perturbation theory, as done in e.g.~\cite{Cruz:2023rmo, Adi:2023qdf, Adi:2023doe}.  A priori, dedicated magnetohydrodynamic (MHD) simulations would be necessary to properly investigate the transition from the linear to the non-linear regime. Even though this is beyond the scope of this project, let us briefly comment on recent advances in this field. In particular, Ref.~\cite{ralegankar2024primordial} showed that the Sheth-Tormen formalism provided a good fit for the  halo mass function in the presence of PMF obtained from hydrodynamical simulations, but also emphasized the strong dependence of the PMF contribution to the relevant damping scale.  They highlighted in particular that dimensionless power spectrum $\Delta_{\rm m}(k)=k^3 P_{\rm m}(k)$\footnote{\repl{$\Delta_{\rm m}(k)$ shall to be confused with $\Delta$ of our Eq.~(\ref{eq:Deltax}), we just use the standard notations.}} gets a   PMF contribution from $k^3 P_\Delta(k)$ that is expected to scale as
$k_A^4\Sigma_B^4 (\lambda_A, z_{\rm rec})$, see discussion around Eq.~(\ref{eq:PDelta}). This implies a contribution from the  fourth power of the damping scale to $\Delta_{\rm m}$. Small theoretical errors in the calculation of the damping scale can thus lead to significant modifications of the amplitude of the PMF contribution to the HMF. Specifically, Ref.~\cite{ralegankar2024primordial} obtained lower values of the PMF-induced HMF contribution in their numerical simulations than from the theoretical estimate, which has been confirmed in \cite{Ralegankar:2024arh}. As will be discussed more in the next sections, the evaluation of bounds from the optical depth to reionization relies strongly on the PMF boost of small-scale structure formation. When evaluating the corresponding constraints in Sec.~\ref{sec:analys_and_results}, we comment on the impact of varying the scaling of the boost amplitude on the bounds.}

\subsection{Imprint on the ionized fraction and IGM temperature}
\label{sec:ADvsDT}
We have implemented the new sources of heating, AD and DT, as well as the magnetic field strength evolution in  {\tt HyRec-2} \cite{Ali-Haimoud:2010hou, Lee:2020obi} to follow the ionized fraction $\overline{x_i}$ and the gas temperature $T_{\rm k}$ as a function of the redshift for $z>35$. At lower redshifts,  {\tt HyRec} does not allow for the self-consistent evolution of the gas temperature and the ionized fraction because it does not account for the formation of stars and galaxies when approaching reionization. We therefore switch to {\tt exo21cmFAST} at $z=35$ to include these effects, and use {\tt HyRec-2} to set the initial conditions for {\tt exo21cmFAST}. {\tt exo21cmFAST} accounts for both exotic energy injection, as described in Sec.~\ref{sec:IGMevol}, and for the modified halo abundance,  as described in Sec.~\ref{sec:growth}, see also Refs.~\cite{Facchinetti:2023slb, Lopez-Honorez:2026bzj, Mosbech:2026hlq, Dandoy:2025twl, Agius:2025nfz, Agius:2025xbj, Lopez-Honorez:2016sur, Decant:2024bpg} for similar applications in the context of dark matter.

\begin{figure}[t]
    \centering
    \includegraphics[width=0.99\linewidth]{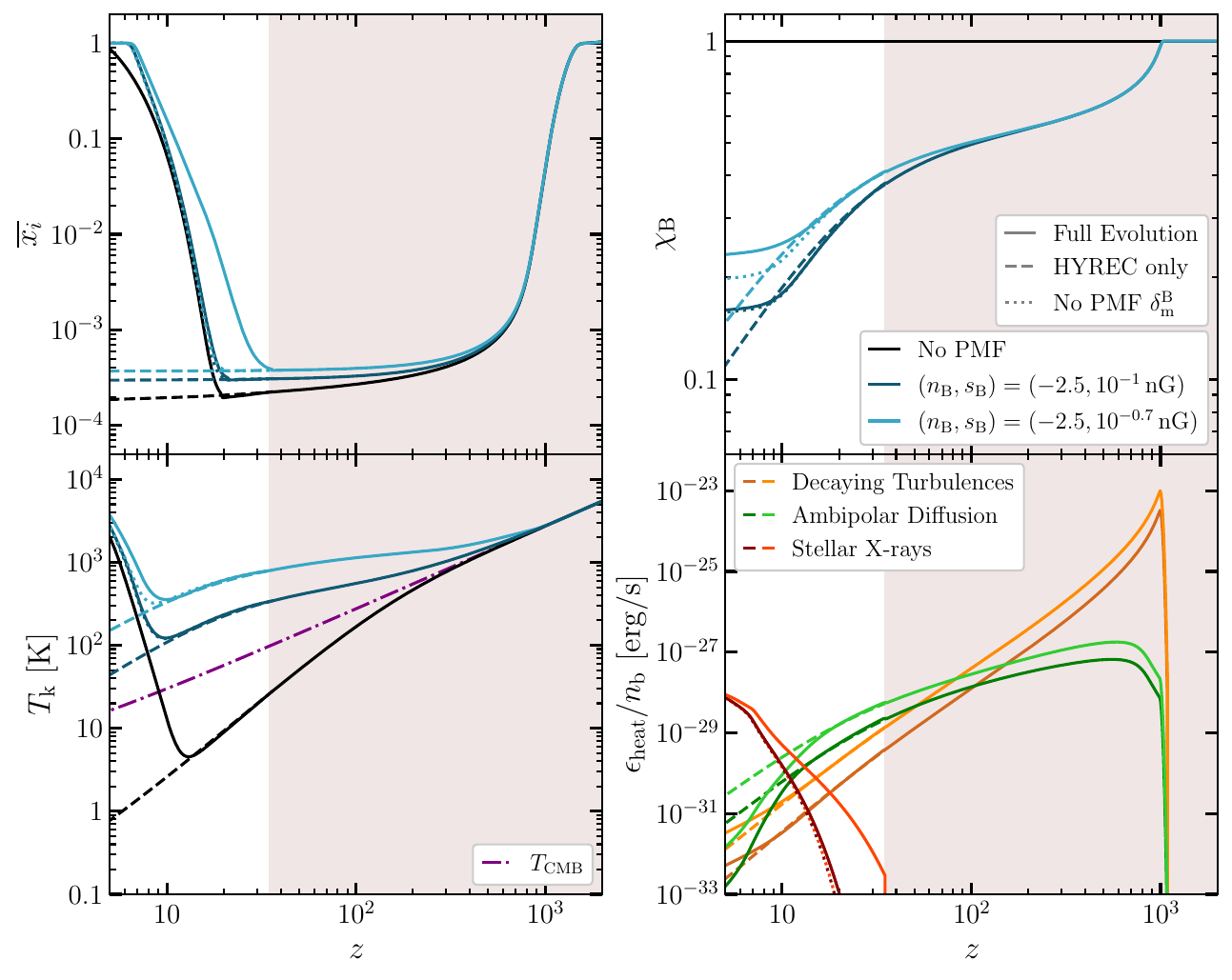}
    \caption{  {\it Left panels:} Ionized fraction (top left) and IGM temperature (bottom left) as a function of the redshift. The black curves correspond to no PMF contribution while the blue curves assume PMF with  $n_{\rm B} = -2.5$  for two values of the amplitude $s_{\rm B} = 10^{-1}\,$nG (darker color lines) and $s_{\rm B} = 10^{-0.7}\,$nG (lighter color lines).  The dashed curves show the output from {\tt HyRec} at $z<35$, while the continuous curves account for the full treatment including both {\tt HyRec} and {\tt exo21cmFAST}. For the curves including PMF contributions, dotted lines neglect the PMF boost of matter perturbation growth. The CMB temperature evolution is shown in purple in the lower plot. {\it Right panels:} corresponding  dimensionless measure of the PMF strength (top right) and energy injection rate into the IGM per baryon (bottom right) as a function of the redshift. In the bottom right plot, the dotted red line also corresponds to the total energy injection rate that is to be expected in the absence of PMF. See the text for more details.
    }
    \label{fig:xe_Tk}
\end{figure}

The left column of Fig.~\ref{fig:xe_Tk} displays the resulting ionized fraction (top left) and IGM temperature (bottom left) evolution as a function of redshift. In the IGM temperature plot, the dot-dashed purple line is the CMB temperature, which simply scales as $(1+z)$. The black curves correspond to the default astrophysical scenario considered here, with the fiducial parameters reported in Tab.~\ref{tab:astro_params}, without any PMF contribution. The output from {\tt HyRec} is shown in dashed lines at $z<35$, whereas the continuous curves show the full evolution using the combination of {\tt HyRec} and {\tt exo21cmFAST}.  The blue curves show how PMFs modify the evolution of the relevant quantities, when we fix $n_{\rm B} = -2.5$ and take two values of the PMF amplitude, $s_{\rm B} = 10^{-1}\,$nG (darker color lines) and $s_{\rm B} = 10^{-0.7}\,$nG (lighter color lines). One can expect the CMB to be sensitive to such PMF parameter values based  on previous analyses of e.g.~\cite{Cruz:2023rmo}.   For the blue curves which include PMF contributions, the dotted lines show the evolution of the quantities when neglecting the PMF boost to matter perturbation growth discussed in Sec.~\ref{sec:growth}. It is visible in the $\overline{x_i}$ plot (top left) that for the larger value of $s_{\rm B}$ ($s_{\rm B}=10^{-0.7}\,$nG, light blue line), this boost plays a major role in the evolution of the late-time ionization history, see also the discussion in~\cite{Cruz:2023rmo}. This is because the increased abundance of halos which host star-forming galaxies leads to more emission of ionizing UV photons. On the other hand, the increase of the ionization floor at $z>35$ is driven by the IGM heating by PMFs, which indirectly affects the ionization history through the temperature-dependence of the recombination/reionization coefficients~\cite{Chluba:2015lpa}. The heating of the medium is clearly visible in the temperature plots, where it shows that for the choice of parameters considered here, the IGM temperature starts to overshoot the CMB temperature once the two temperatures decouple.

The top right panel of Fig.~\ref{fig:xe_Tk} shows the redshift evolution of the dimensionless amplitude of the magnetic fields, where the solid black line corresponds to the case of no PMF decay. The lower right panel shows the IGM heating rate per baryon for each of the three sources of IGM heating, namely stellar X-rays (red curves), PMF decaying turbulence (orange curves) and ambipolar diffusion (green curves). Lighter (darker) color curves correspond again to the larger (smaller) value of $s_{\rm B}$; dashed lines show the evolution from {\tt HyRec}, which neglects star formation and reionization; dotted lines neglect the PMF boost to matter perturbation growth while the continuous lines include all effects.  
From the top right panel, it is clear that the PMF energy losses lead to a significant decrease of the PMF amplitude between recombination and reionization, up to $\sim$ one order of magnitude for the choice of parameters made here. This energy loss is significant in the entire parameter space we consider, other choices of $(n_{\rm B}, s_{\rm B})$ can lead to even larger decreases in $\chi_{\rm B}$. This is in contrast to the approach followed in most of previous analyses, which have neglected such energy losses, thus effectively setting $\chi_{\rm B}(z)=1$ (corresponding to the black line in the upper right panel). We can  therefore expect that such analyses have found more dramatic effects of PMFs on the ionized fraction and IGM temperature for the same set of PMF parameters than we do.
 
The energy losses are driven by PMF DT at early times, as visible with the orange curves in the lower panel, see also~\cite{Sethi:2004pe, Chluba:2015lpa}. For this specific model, the AD losses start dominating at $z\lesssim 100$. Comparing dashed to continuous lines, we see that reionization leads to a decrease of such losses at the lowest redshifts  and to a saturation of $\chi_{\rm B}$ at the lowest redshift. This is due to the  $(1-x_\e)/x_\e$ dependence of the AD losses in Eq.~(\ref{eq:epsheatAD}). On the other hand, X-ray sources  dominate IGM heating at $z\lesssim10$ (red curves), the abundance of which gets significantly enhanced by the extra growth of matter perturbations induced by PMF, especially for larger values of $s_{\rm B}$. Let us emphasize that both the dark and light red curves in the bottom right panel show the X-ray energy injection rate given the same astrophysics parameters. The light red curve displays an enhanced rate due to the boost of small scale structures induced by higher PMF amplitude $s_{\rm B}$, not higher X-ray luminosity $L_X$. This is visible when comparing the dotted red curves (which lie on the top of each other for the two $s_{\rm B}$ values) to the continuous ones. Altogether, our illustrative example shows that  PMF energy injection  into heating together with their boost of small scale structures can induce an earlier reionization and increase the IGM temperature by several orders of magnitude (blue curves in the top and bottom left panels) with respect to the no-PMF case (black curves).

\section{Optical depth and UV luminosity function constraints}
\label{sec:analys_and_results}

Because PMFs are expected to modify both the IGM history and structure formation, they can be constrained by available late-time cosmological data, associated to reionization and  UV luminosity observations (see~\cite{Sethi:2004pe, Kunze:2013uja, Chluba:2015lpa, Paoletti:2018uic, Bera:2020jsg, Jedamzik:2023csc, Jedamzik:2023rfd, Cruz:2023rmo, Schleicher:2011jj, Katz:2021iou} and discussion in Sec.~\ref{sec:ADvsDT}).\footnote{Lyman-$\alpha$ data constraints have recently been studied in~\cite{Pavicevic:2025gqi}. To account for those self-consistently would imply a careful modeling of the signal that is beyond the scope of this paper. } They are sensitive to the presence of PMFs through modifications of  the evolution of the ionized fraction and/or the clumping properties of matter at small scales. Those effects are potentially degenerate with astrophysical parameters modeling the star formation history as well as the X-ray and UV fluxes from stars.

Here we consider a reionization history and UV luminosity flux evaluated with \texttt{exo21cmFAST} that  depend on  the astrophysical parameters listed in Tab.~\ref{tab:astro_params}. Several parameters were already introduced in Sec.~\ref{sec:growth}.
The stellar mass fraction and fraction of ionizing UV photons that escape from a galaxy hosted in a halo of mass $M$ are modeled as~\cite{Park:2018ljd, Qin:2020xyh, Stefanon_2021, Shuntov:2022qwu} 
\begin{eqnarray}
    f_\star &=&   f_{\rm b}\,  {\rm min}\left\{1, f_{\star, 10} \left(\frac{M}{10^{10} ~{\rm M}_\odot}\right)^{\alpha_\star}\right\}
\label{eq:fstar}\\
   f_{\rm esc} &=&  {\rm min}\left\{1, f_{{\rm esc} , 10} \left(\frac{M}{10^{10} ~{\rm M}_\odot}\right)^{\alpha_{\rm esc}}\right\}
   \label{eq:fesc}
\end{eqnarray}
respectively, where $f_{\star, 10}$ ($f_{{\rm esc}, 10}$) is the stellar mass fraction (UV photon escape fraction) of a halo with mass $10^{10}\,{\rm M}_\odot$ and $\alpha_\star$ ($\alpha_{\rm esc}$) captures the dependency on the halo mass. The stellar mass fraction is defined as $f_\star=M_\star/M$ while the
 star formation rate is given by $\dot M_\star=M_\star H(z)/t_\star $,
where  $t_\star$ is the dimensionless parameter describing the typical star formation timescale in units of the Hubble time.

In the following, we detail our methodology to obtain constraints on PMFs from reionization and UV luminosity data, as well as our results. The entire pipeline is similar to that of \cite{Facchinetti:2025hou} and is based on the emulator {\tt NNERO} for the ionized fraction during the EoR and the optical depth to reionization. We also briefly discuss how the emulator works and discuss the training dataset used for this work before addressing the MCMC analysis.

\begin{table}[t]
    \centering
    \begin{tabular}{|c|c|c|c|c|c|}
    \hline &&&&&\\[-1em]
      & {\bf \small Param.} & {\bf \small Units} & {\bf \small Range} & {\bf \small Flat prior} & {\bf \small Fiducial}\\[3pt]
    \hline &&&&&\\[-1em]
    \multirow{3}{*}{Star formation rate} & $f_{\star, 10}$ & $-$  & $[10^{-2.5}, 1.0]$ & $\log_{10}$ & $10^{-1.2}$\\[3pt]
    & $\alpha_\star$ & $-$  & $[-0.1, 1.0]$ & linear & $0.5$ \\[3pt]
    & $t_\star$ & $-$ & $[0.0, 1.0]$ & linear & $0.5$\\[3pt]
    \hline &&&&&\\[-1em]
    \multirow{2}{*}{UV escape fraction} & $f_{{\rm esc}, 10}$ & $-$ & $[10^{-2.5}, 1.0]$ & $\log_{10}$ & $10^{-1.0}$ \\[3pt]
    & $\alpha_{\rm esc}$ & $-$ & $[-1.0, 0.5]$ & linear & $0.0$ \\[3pt]
    \hline &&&&&\\[-1em]
    \multirow{2}{*}{X-ray emission} & $L_\X$ & ${\rm erg\,yr\,s}^{-1}{\rm M}_\odot^{-1}$ & $[10^{38}, 10^{42}]$ & $\log_{10}$ & $10^{40.5}$ \\[3pt]
    & $E_0$ & keV & [0.1, 1.5] & linear & $0.5$  \\[3pt]
    \hline &&&&&\\[-1em]
    Star formation efficiency & $M_{\rm turn}$ & M$_\odot$ & $[10^{6}, 10^{10}]$ & $\log_{10}$ & $10^{8.6}$\\[3pt]   
    \hline
    \end{tabular}
    \caption{Astrophysical parameters of our model and the range of values used to generate the training sample of {\tt NNERO} and define the MCMC priors. The fiducial values in the rightmost column are used for the $21\,$cm forecast and for illustrative purposes in the plots.
    }
    \label{tab:astro_params}
\end{table}

\subsection{Reionization observational constraints}
\label{sec:an-reio}

PMF are known to leave different imprints in the CMB inducing spectral distortions~\cite{Jedamzik:1999bm, Zizzo:2005az, Kunze:2013uja, Saga:2017wwr}, anisotropic expansion~\cite{Barrow:1997mj}, CMB polarization, non-gaussianities, inhomogeneous recombination~\cite{Jedamzik:2018itu} and affecting CMB anisotropies~\cite{Subramanian:1998fn,2002MNRAS.335L..57S,Mack:2001gc,Lewis:2004kg,Kahniashvili:2005xe,Chen:2004nf,Lewis:2004ef,Tashiro:2005hc, Yamazaki:2006bq,Giovannini_2006,Paoletti:2010rx,2012PhRvD..86d3510S, Kunze:2013iwa,Kunze:2011bp,Kunze:2010ys,Paoletti:2012bb,Ade:2015xua,Sethi:2004pe,Chluba:2015lpa,Paoletti:2018uic, Seshadri:2000ky} 
and reionization~\cite{Sethi:2004pe,Paoletti:2022gsn,Cruz:2023rmo,Seshadri:2000ky,Schleicher:2011jj,Pandey:2014vga}. Typically, those CMB analyses constrain PMFs with $n_{\rm B}\simeq -3$  (sometimes referred to as scale-invariant PMF) to have a strength today  $\sigma_{\rm B,0}\lesssim 0.1-0.05\,$nG over a coherence scale of $1\,$Mpc (see~\cite{Jedamzik:2018itu} for a summary).  

Here we revisit the PMFs' impact on the optical depth to reionization, $\tau$, due to exotic heating and the boost of structures, taking into account the energy losses in their evolution. We evaluate $\tau$ as
\begin{equation}
    \tau=\sigma_{\rm T}\int_{0}^{ z_{\rm min}}\dd z \, \overline{n_i} \frac{\dd l}{\dd z},
    \label{eq:opt}
\end{equation}
where $\sigma_{\rm T}$ is the Thompson cross-section, $\dd l/\dd z$ is the line-of-sight proper distance per unit redshift, $\overline{n_i}$ is the free electron number density averaged over sky directions computed from $\overline{x_i}$ of Eq.~(\ref{eq:xHIIdb}) and $z_{\rm min}$ is the redshift at which the free electron fraction displays a minimum within the simulated redshift range, see~\cite{Facchinetti:2025hou} for details. This methodology follows from~\cite{Poulin:2016anj}  and   is implemented in the public Boltzmann code {\tt CLASS}~\cite{Lesgourgues:2011re} to  deal with e.g.~exotic energy injection analyses.
The modification to $\tau$ becomes especially significant when the additional small-scale power induced in the matter power spectrum by PMFs is sufficient to trigger earlier star formation, leading to earlier reionization, as illustrated in Fig.~\ref{fig:xe_Tk} in the $s_{\rm B}=10^{-0.7}\,$nG case, and an enhanced optical depth to reionization~\cite{Cruz:2023rmo}. In standard CMB analyses~\cite{Planck:2018vyg}, the optical depth to reionization is treated as a free parameter of the fitted $\Lambda$CDM model assuming a $\tanh$ shape for the ionized fraction during the EoR. However, given an astrophysical framework like the one considered in \texttt{exo21cmFAST}, the optical depth can be computed self-consistently from models of star formation and the evolution of the IGM that shape the reionization history. As an additional constraint, strong evidence for reionization to be completed (or nearly completed) by $z \approx 6$ has been found in e.g.~\cite{McGreer:2014qwa} from the spectroscopy of bright quasars studying the fraction of Ly-$\alpha$ and Ly-$\beta$ forest regions with zero transmitted flux. This imposes lower bounds on the ionized fraction at $z=5.6$ and $5.9$, see the discussion in Sec.~\ref{sec:MCMC}.

\begin{figure}[t]
    \centering
   \includegraphics[width=0.51\linewidth]{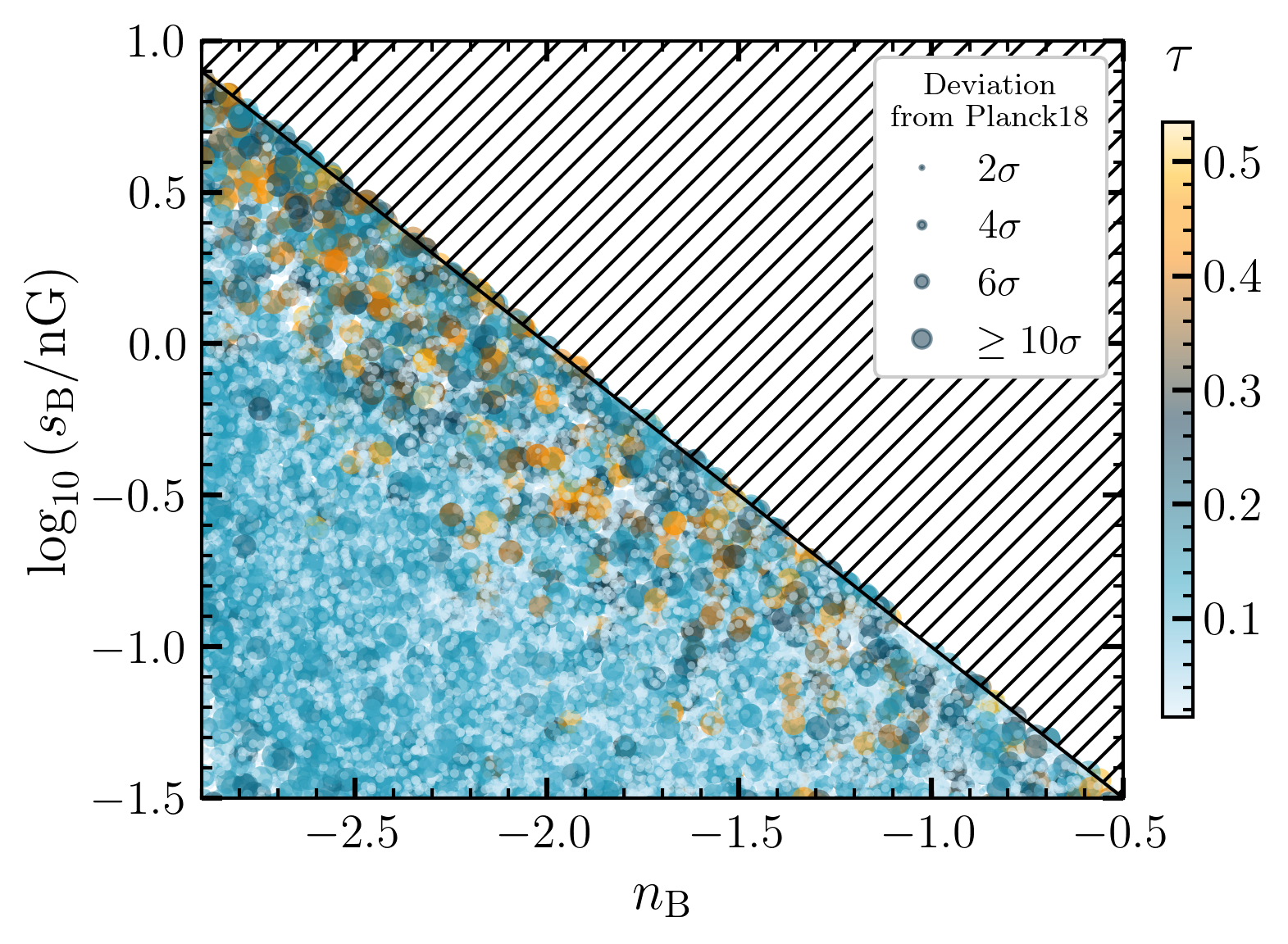} 
   \includegraphics[width=0.47\linewidth]{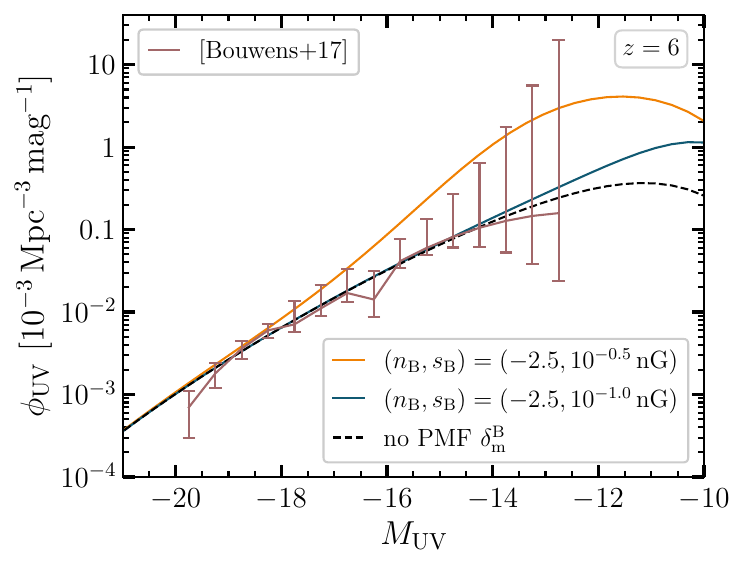}
    \caption{{\bf Left panel.} Distribution of the optical depths to reionization simulated for the training, validation and test of {\tt NNERO} in the plane $(n_{\rm B}, s_{\rm B})$. Colors indicate the value of the optical depth to reionization $\tau$ and the size of the marker depends on the deviation from the Planck18 value. The hatched upper right region is not considered in our analysis. {\bf Right panel.} Impact of the increased matter clustering due to primordial magnetic fields on the UV luminosity function at redshift 6. The UV LF measurements from Tab.~5 of~\cite{Bouwens_2017} are shown in brown with $1\,\sigma$ error bars. The black dashed line shows $\phi_{\rm UV}$ for the fiducial parameters of Tab.~\ref{tab:astro_params}, assuming only CDM clustering, while the colored lines show the effect of PMFs for two different sets of parameters.}
    \label{fig:UVLF_data}
\end{figure}

For illustration, we show in the left panel of Fig.~\ref{fig:UVLF_data} a distribution of sampled points in the plane  $(n_{\rm B}, s_\B)$ for random values of the astrophysical parameters. The colors of the markers show the value of optical depth to reionization $\tau$ while their size tracks the deviations from the value reported by Planck 2018~\cite{Planck:2018vyg} (TTTEEE+lowE+lensing), $\tau_{\rm Planck}=0.0544\pm 0.0073$. It can be seen that for fixed $n_{\rm B}$, larger values of $s_{\rm B}$ induce stronger deviations above  $\tau_{\rm Planck}$. The hatched upper right region is not considered in our analysis as it modifies the reionization history in ways that are completely excluded by Planck and because focusing on the lower region improves the performance of the emulator. Large deviations can also appear for small $s_{\rm B}$ for certain combinations of astrophysical parameters. For instance, a small value of the threshold mass for star-forming galaxies $M_{\rm turn}$ combined with a low value of $\alpha_\star$ results in a high star formation rate at early times inside small host halos, which will yield early reionization and a too large optical depth to reionization even without PMFs.

\subsection{UV luminosity constraints}
\label{sec:an-UV}

In addition to the increased ionized fraction discussed above, we also expect PMFs to impact galaxy UV luminosity functions, see also~\cite{Schleicher:2011jj}. The UV LF, $\phi_{\rm UV}$, measures the number density of objects with a given UV luminosity per UV luminosity interval, in terms of the UV magnitude, $M_{\rm UV}$. We evaluate the UV LFs from the HMF as \cite{Park:2018ljd}
\begin{equation}
    \phi_{\rm UV}(M_{\rm UV}) =f_{\rm duty} \frac{{\rm d}n}{{\rm d}M} \left|\frac{{\rm d}M}{{\rm d}M_{\rm UV}}\right|\,.
    \label{eq:UVLF}
\end{equation}
The halo mass and the UV magnitude are related through the star formation rate \cite{10.1093/mnras/stw980, Oke:1983nt} as
\begin{equation}
    \frac{\dot M_\star(M)}{10^{10}\, {\rm M}_\odot} = \Gamma_{\rm UV} 10^{-0.4 M_{\rm UV}} \,,
    \label{eq:Mdot_vs_MUV}
\end{equation}
with $\Gamma_{\rm UV} = 5.16 \times 10^{-18} \,{\rm yr}^{-1}$. 

The additional PMF-induced matter clustering on small scales discussed in Sec.~\ref{sec:growth} affects the UV LFs through an increase of the HMF over a range of halo masses set by the Jeans and the Alfven scales, see also the discussion in~\cite{Schleicher:2008aa}.

For illustration, we show in the right panel of Fig.~\ref{fig:UVLF_data} the UV luminosity function as a function of the UV magnitude at redshift $z = 6$ for different parameter choices. The brown curve with vertical error bars shows the data points with $1\,\sigma$ error bars, taken from the Tab.~5 of Ref.~\cite{Bouwens_2017}, who studied a sample of galaxies at $z \sim 6$ inside Hubble Frontier Fields clusters.
The dashed black line represents the expectation assuming no clustering effect of PMFs. 
The colored curves show the impact of PMFs, given the same astrophysical model but considering two values of $s_{\rm B}$ for a fixed $n_{\rm B}=-2.5$. From the discussion in Sec.~\ref{sec:growth}, we expect that increasing values of $s_{\rm B}$ induce an enhancement of the HMF at increasing halo masses, or decreasing values of $M_{\rm UV}$, as can be observed in Fig.~\ref{fig:UVLF_data}. We thus derive our constraints using UV LF data from Refs.~\cite{Bouwens_2017, Bouwens_2021}.\footnote{New UV luminosity function measurements from JWST are now available. For the purpose of this work, the data from the faint end of the luminosity functions is already constraining the PMF scenarios, a regime in which both HST and JWST data are consistent.   Incorporating the JWST data would necessitate modifications to the {\tt 21cmFAST} stellar model to reproduce the bright end of the luminosity functions,  see also e.g.~\cite{Dhandha:2025gib}.}

\subsection{Emulation of the optical depth with {\tt NNERO}}
\label{sec:opdep}

As a single simulation of the optical depth with {\tt 21cmFAST} is costly in terms of CPU time, the emulator \texttt{NNERO} was introduced in Ref.~\cite{Facchinetti:2025hou} to allow for the consistent joint analysis of late-time probes with CMB data and was coupled to \texttt{MontePython}~\cite{Audren:2012wb, Brinckmann:2018cvx} and \texttt{CLASS} to perform an analysis over astrophysical parameters and $\Lambda$CDM parameters (excluding $\tau$) simultaneously. It was nevertheless shown that performing the inference over the full parameter space leads to very similar results as when fixing the cosmological parameters and collapsing the CMB likelihood into a likelihood for the optical depth to reionization. For simplicity, we therefore adopt the latter approach in this work. Further details on the likelihood are provided in the next subsection. 

\texttt{NNERO} emulates the evolution of the free-electron fraction during the EoR by learning from a large set of simulated ionization histories where all astrophysical parameters listed in Tab.~\ref{tab:astro_params} are varied within the reported ranges. Then, the PMF parameters $s_{\rm B}$ and $n_{\rm B}$ are  varied logarithmically and linearly, respectively, within the ranges
\begin{equation}
s_{\rm B}/{\rm nG} \in [10^{-5},10] \quad \text{and} \quad n_{\rm B} \in [-2.9,-0.5]\,,
\end{equation}
subject to the additional condition
\begin{equation}
\log_{10} (s_{\rm B}/{\rm nG}) \le -n_{\rm B} - 2  \,.
\end{equation}
This upper cut on $s_{\rm B}$ improves the emulator’s precision in the phenomenologically relevant region of parameter space as larger values will be excluded by data after marginalization over the range of astrophysics parameters, see also Sec.~\ref{sec:resultsEoR+Planck-τ+UV}.

Within \texttt{NNERO}, the input parameters are first processed by a lightweight neural network classifier that implements a {\it selection cut}. This classifier identifies whether a given parameter combination produces an ionization history that is not already excluded at more than $5\,\sigma$ by the ionized fraction upper bounds at $z\sim 6$ of~\cite{McGreer:2014qwa} and that does not violate the Planck constraint on the optical depth at more than $10\,\sigma$. Parameter sets passing this selection are then fed into a deep neural network regressor, which predicts the principal component coefficients of $x_\e(z)$. This two-stage pipeline allows the regressor to focus exclusively on physically viable ionization histories, thereby improving its performance.
In total, 105,044 simulations were generated, of which 44,229 passed the selection cut. For the classifier, 84,036 samples were used for training, 10,504 for validation, and 10,504 for testing. Among the accepted samples, the regressor was trained on 35,382 histories, validated on 4,463, and tested on 4,384. After training, the accuracy of the classifier is evaluated as $98\%$ on the test set. To quantify the goodness of the emulator, we consider the error
\begin{equation}
    \epsilon_\tau \equiv \frac{|\tau_{\rm pred} - \tau_{\rm true}|}{\tau_{\rm true}}\,,
\end{equation}
where $\tau_{\rm true}$ is an optical depth value of the test set, given by the simulation,  and $\tau_{\rm pred}$ is the associated value predicted by {\tt NNERO}. After training, the average and spread of $\epsilon_\tau$ are 0.07\% and 1.1\% respectively. These are well within the required precision for this analysis.

In the next subsections, we present our results in terms of the $(s_{\rm B}, n_{\rm B})$ parameter space. However, in order to ease comparisons with previous literature and the PMF parameter $\sigma_{\rm B, 0}=s_{\rm B}\chi_{\rm B}(0)$ used therein, we implement a third neural network that learns $\chi_{\rm B}(0)$ from the input parameters.

\subsection{MCMC analysis}
\label{sec:MCMC}

We use the following likelihoods in our analysis:
\begin{itemize}
\item {\bf Planck-$\tau$}. This likelihood corresponds to the posterior constraint on $\tau$ from Planck 2018 TTTEEE+lowE+lensing data \cite{Planck:2018vyg}. It is chosen as a Gaussian with mean $\mu_\tau =  0.0544$ and standard deviation $\sigma_\tau = 0.0073$ and evaluated by the {\tt NNERO} regressor.

    \item {\bf EoR.} This likelihood is always included in order to be consistent with the selection cut operated by the classifier in {\tt NNERO}. It is based on the free electron fraction measurements presented in Ref.~\cite{McGreer:2014qwa}. It is implemented as a  asymmetrical Gaussian distribution imposing $\overline{x_i}(z=5.6) = 0.96_{-0.05}^{+\infty}$ and $\overline{x_i}(z=5.9) = 0.94_{-0.05}^{+\infty}$ at the $1\,\sigma$ confidence level -- where the $+\infty$ upper notation simply indicate that these are lower bound on the value of the ionized fraction.\footnote{While not critical for this study, note that more recent studies on the fraction measurements have been performed, see e.g.~\cite{Qin:2024cna}.}

    \item {\bf UV LFs.} This likelihood refers to UV luminosity functions data from the Hubble Space Telescope. Following Ref.~\cite{HERA:2021noe} it is taken as a split Gaussian, evaluated using the data from Refs.~\cite{Bouwens_2017, Bouwens_2021}, truncated to a magnitude $M_{\rm UV} > -20$. 
\end{itemize}

We perform three independent MCMC analyses using the sampler {\tt emcee}~\cite{2013PASP..125..306F}, combining the likelihoods as follows: (i) {\bf EoR+Planck-$\tau$}, (ii) {\bf UV LFs}, and (iii) {\bf EoR+Planck-$\tau$+UV LFs}. In the first and third configurations we vary all 10 parameters listed in Tab.~\ref{tab:astro_params}. As the UV LFs are independent of the parameters related to the escape fraction and X-ray emission, the parameter space is reduced from 10 to 6 dimensions in the second setup. Convergence is assumed to be reached when the chain length exceeds 20 times the autocorrelation length.

\subsection{Results}
\label{sec:resultsEoR+Planck-τ+UV}

\begin{figure}[t]
    \centering

    \includegraphics[width=0.49\linewidth]{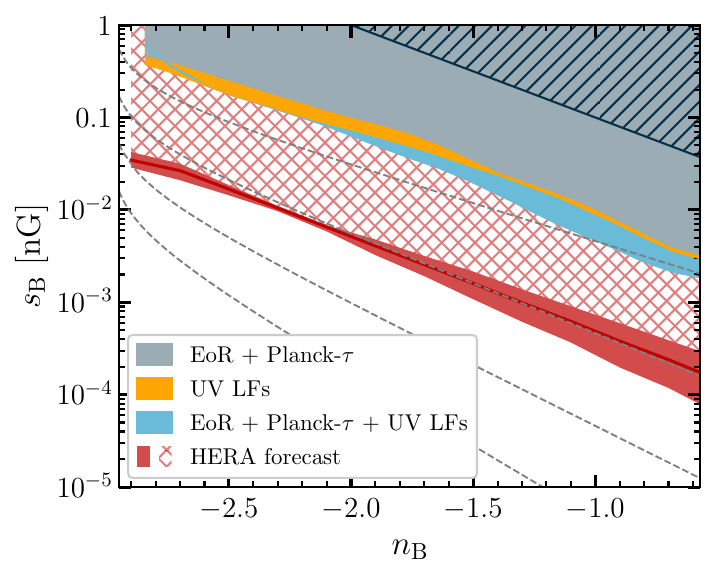}
    \includegraphics[width=0.49\linewidth]{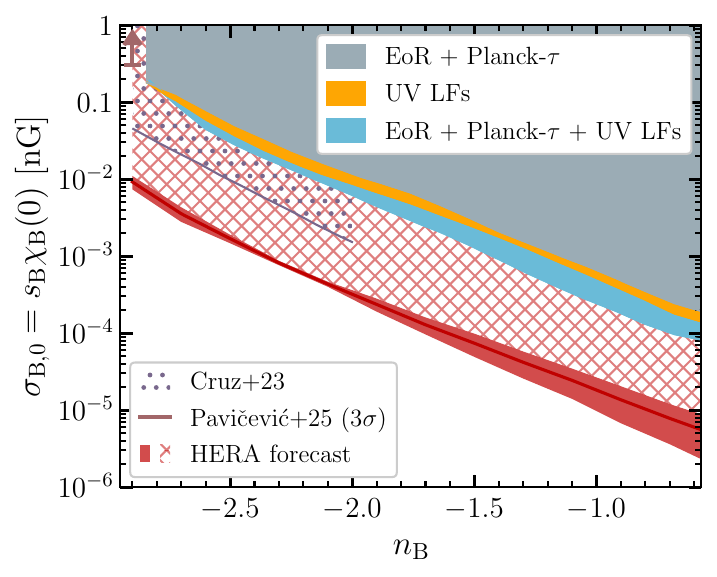}
    \caption{95\% CL constraints on $s_{B}$ (left panel) and $\sigma_{B, 0}$ (right panel) in terms of $n_{\rm B}$. The orange region is obtained from UV LF likelihood only. The gray region is obtained from Planck-$\tau$ and the EoR constraint on $x_\e$.  The blue shaded region is the combined constraint. The crossed red region is the forecast for the HERA $21\,$cm telescope array under assumption of the fiducial parameters listed in Tab.~\ref{tab:astro_params}. The uncertainty on the forecast bound is shown with the red shaded band around the lower limit, see  Sec.~\ref{sec:forecast} for details. In the left panel, the dashed gray lines are contours of constant Lorentz force. In the right panel, we report the allowed region by CMB data by Ref.~\cite{Cruz:2023rmo} (labeled as Cruz+23) with a dotted purple area. The brown arrow shows the $3\,\sigma$ upper limit from Ref.~\cite{Pavicevic:2025gqi} (labeled as Pavi\v{c}evi\'{c}+25) for $n_\B=-2.9$.}
    \label{fig:constraints}
\end{figure}

Figure~\ref{fig:constraints} projects the exclusion regions at 95\% confidence level (CL) for the EoR+Planck-$\tau$ (gray), UV LFs (orange), and EoR+Planck-$\tau$+UV LFs (blue) analyses in the ($n_\B,s_{\rm B}$) plane on the left panel and the ($n_{\rm B}, \sigma_{B, 0}$) plane on the right panel. The shaded region in the top right of the left plot shows the parameter space where the emulator has not been trained, as it would anyway be completely excluded by the data. We see  that the addition  of UV luminosity data w.r.t. e.g. the analysis of~\cite{Cruz:2023rmo} allows to improve on the limit extracted from just data on the reionization history alone (dark and light gray contours). It is visible that the upper right hatched region, which was not considered in our MCMC analysis to improve the emulator performance, lies well within the region excluded by data. 

Interestingly, we note that both reionization (EoR+Planck-$\tau$) and UV LF constraints display a similar scaling in the $(n_{\rm B},s_{\rm B})$ plane. They are thus expected to originate from the same underlying effect: the enhancement of small-scale power in the matter power spectrum. The latter is due to the Lorentz force driven term $\propto {\bf S}_L$ in the baryons Euler Eq.~(\ref{eq:sourceEulerb}). As discussed in Sec.~\ref{sec:analys_and_results}, when the additional small-scale power is sufficiently large, it promotes earlier and more efficient star formation, simultaneously increasing the UV luminosity functions and inducing an earlier onset of reionization that translates into an increase of the optical depth to reionization. Heating alone has only a marginal impact on our constraint on the optical depth to reionization. To highlight this common physical origin, dashed  gray lines contours of fixed values of the squared Lorentz force average are shown in the left panel of Fig.~\ref{fig:constraints}. Those contours  were obtained setting to a constant~Eq.~(\ref{eq:Lforcesq}), which scales as $\langle |{\bf S}_L|^2\rangle$,  and using Eq.~(\ref{eq:sigma_B}). They thus correspond to lines satisfying the relation
\begin{equation}
    \log_{10} \left(\frac{s_{\rm B}}{\rm nG}\right) = -\frac{5+n_{\rm B}}{4} \left\{C +\log_{10}\left[f_L(3+n_{\rm B})\right]\right\}
    \label{eq:AD_const_Lorentz}
\end{equation}
for constant values of $C$, where we set $\chi_{\rm B}$ to 1 for simplicity. The dashed gray lines shown in the left panel of Fig.~\ref{fig:constraints} correspond to $C\in\{3,4,5,6\}$. We observe that both the lower edges of the derived constraints (the bottom of the colored regions) and the contours of constant Lorentz force (dashed gray lines) exhibit very similar scaling.

To ease the comparison with previous literature, we reinterpret the obtained limits in terms of the more commonly used parameter $\sigma_{{\rm B},0}$ (instead of $s_{\rm B}$) in the right panel, using the value of $\chi_{\rm B}(0)$ output by the emulator. 

In the right panel, we also report as the dotted purple area the exclusion region obtained by~\cite{Cruz:2023rmo} by imposing temperature and polarization data constraints on the optical depth to reionization (they use $\tau = 0.0627 \pm 0.0050$ at $2\,\sigma$). Despite the fact that we use an extra source of data to bound the magnetic field parameters, our 95\% CL constraints appear to be looser than theirs by roughly one order of magnitude for the same values of $n_{\rm B}$. This relaxation of constraints is mainly due to the treatment of energy losses of the magnetic field over time, which we track through the evolution of $\chi_{\rm B}$ in Eq.~(\ref{eq:chiBdot}). As shown in Fig.~\ref{fig:xe_Tk}, this decrease of the PMF field strength is non-negligible. Our analysis further differs from that of~\cite{Cruz:2023rmo} in the methodology to extract the limit, as we perform a full MCMC analysis rather than a simplified Fisher matrix forecast. Finally, we overlay on the same plot the Lyman-$\alpha$ forest constraint obtained in \cite{Pavicevic:2025gqi} for $n_{\rm B} = -2.9$ with a brown arrow, noting that this result is reported at the $3\,\sigma$ level (and not at 95\% CL). Nonetheless, our constraint is of the same order of magnitude. See also \cite{Pandey:2012ss} for an earlier analysis based on Lyman-$\alpha$ data.

\repl{
As mentioned in Sec.~\ref{sec:HMF}, MHD simulations allow for a more precise evaluation of the transition from linear to non-linear evolution of the perturbations. Refs.~\cite{ralegankar2024primordial, Ralegankar:2024arh} performed such simulations (neglecting energy injection) and concluded that the amplitude of the PMF-induced boost could be significantly lower than the simple estimate made here based on linear perturbation evolution. In the light of those results, we have explicitly checked that a reduction of the amplitude of the PMF-induced boost to the matter power spectrum by three orders of magnitude degrades our CMB and UV LF constraints by one order of magnitude in $s_{\rm B}$ for fixed $n_{\rm B}$. The $21\,$cm forecast derived in the next section will however be unaffected.
}

\begin{figure}[t]
    \centering
    \includegraphics[width=0.99\linewidth]{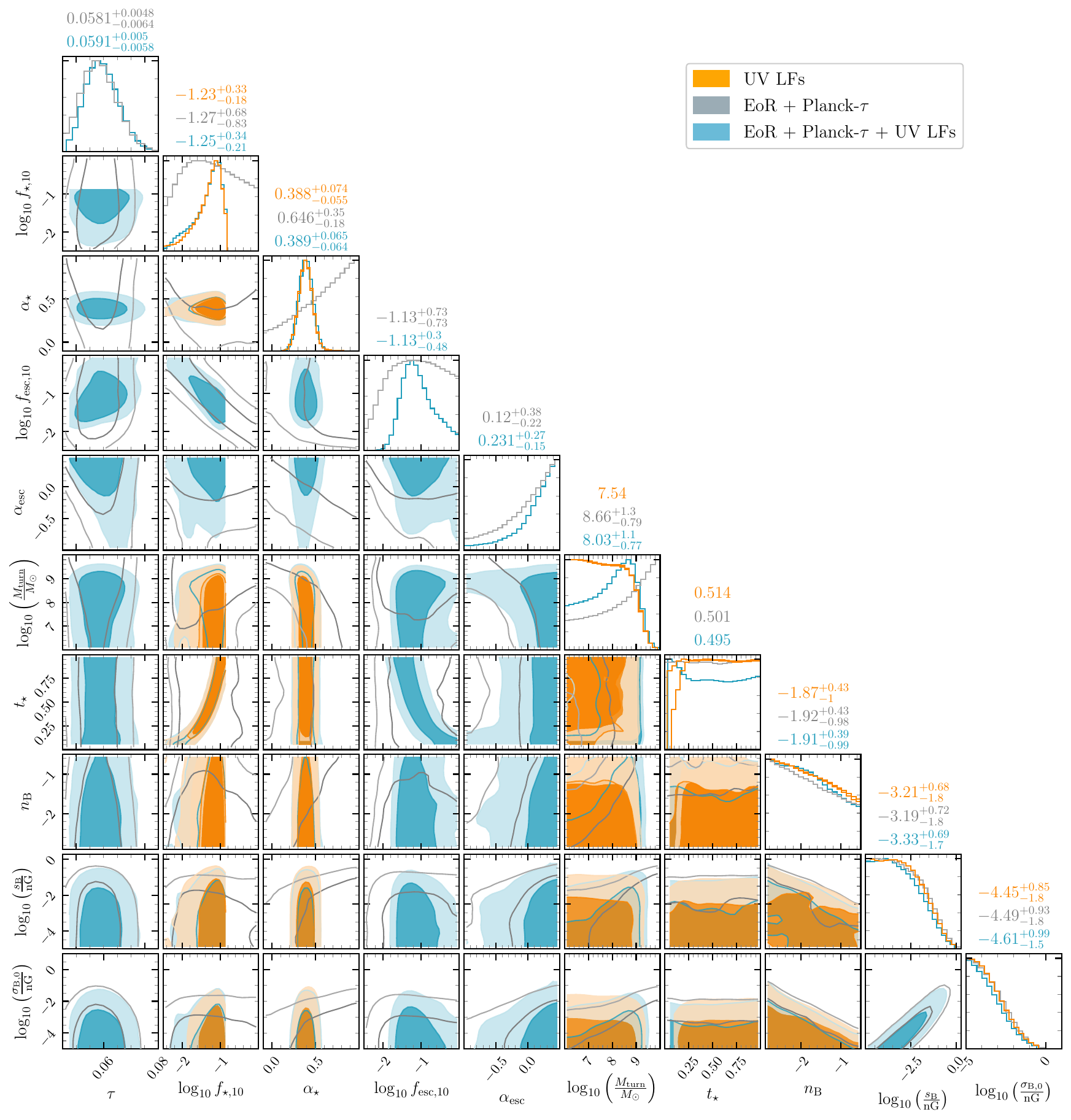}
    \caption{Corner plot of the 1D posterior distribution  at 95\% CL and 2D posteriors at \ 68\% (darker shades) and 95\% CL (lighter shades) for all three likelihood combinations. For better visibility, the parameters $L_{\rm X}$ and $E_0$, which are largely unconstrained by the data, are omitted. Orange, gray and blue contours correspond to posteriors for UV LF data only,  EoR only and the combined dataset respectively. }
    \label{fig:triangle}
\end{figure}

We provide in Fig.~\ref{fig:triangle}  all the projections of marginalized posterior contours at 68\% and 95\% CL (darker and lighter colors respectively) of the  EoR+Planck-$\tau$ (gray),   UV LFs (orange), and EoR+Planck-$\tau$+UV LFs (blue) analyses. As the X-ray parameters $L_{\rm X}$ and $E_0$ do not sizably affect any of the observables and remain entirely unconstrained in this configuration, we omit them from the plot for clarity. The off-diagonal (diagonal) panels show the 2D (1D) posteriors marginalized over all other parameters. UV LFs are only sensitive to star formation-related parameters, but strongly constrain $f_{\star,10}$ and $\alpha_\star$, as already observed in e.g.~\cite{HERA:2021noe}.  No strong correlation between the PMFs and any particular astrophysical parameter can be observed.

\section{21$\,$cm forecast}
\label{sec:21cm}

Spin-flip transitions between the two ground state energy levels of neutral hydrogen give rise to the absorption or emission of photons with wavelength $21\,$cm. As neutral hydrogen is the dominant constituent of baryonic matter between recombination and  reionization, it is expected that the measurement of the intensity of the $21\,$cm signal as a function of time and space will allow to infer the IGM history during Cosmic Dawn and the Epoch of Reionization with high precision~\cite{Pritchard_2012}. It has already been shown that earth-bound radio telescope arrays, which probe the IGM fluctuations at redshifts $z\sim 5-20$, can be sensitive to exotic energy injection and modifications of small-scale structure formation, see e.g.~\cite{Evoli:2014pva, Lopez-Honorez:2016sur, Facchinetti:2023slb, Agius:2025xbj, Agius:2025nfz, Bhaumik:2024efz, Bhaumik:2026mlc, Katz:2021iou, Cruz:2023rmo,Dandoy:2025twl}. In this section, we briefly comment on the possibility for the Hydrogen Epoch Reionization Array (HERA) telescope~\cite{HERA:2021bsv,HERA:2021noe,HERA:2025ajm}, currently under deployment in South Africa,   to further  constrain PMFs. We evaluate its sensitivity to PMFs by simulating the $21\,$cm signal with {\tt exo21cmFAST} and performing a simple Fisher forecast assuming a single population of galaxies modeled with the fiducial astrophysical parameters tabulated in Tab.~\ref{tab:astro_params}. Previous works~\cite{Cruz:2023rmo, Bhaumik:2024efz, Bhaumik:2026mlc} have already provided $21\,$cm cosmology forecasts for  PMFs; our approach followed here improves on~\cite{Cruz:2023rmo} as we do not neglect PMF energy losses and also include AD and DT heating effects, and on~\cite{Bhaumik:2024efz,Bhaumik:2026mlc} in the more advanced modeling of the $21\,$cm signal. Furthermore,  we discuss the dependence of our 21$\,$cm forecasts to the choice of astrophysical parameters.

\subsection{PMF imprint on the 21$\,$cm signal}
\label{sec:21cmimprint}

\begin{figure}[t]
    \centering
    \includegraphics[width=0.6\linewidth]{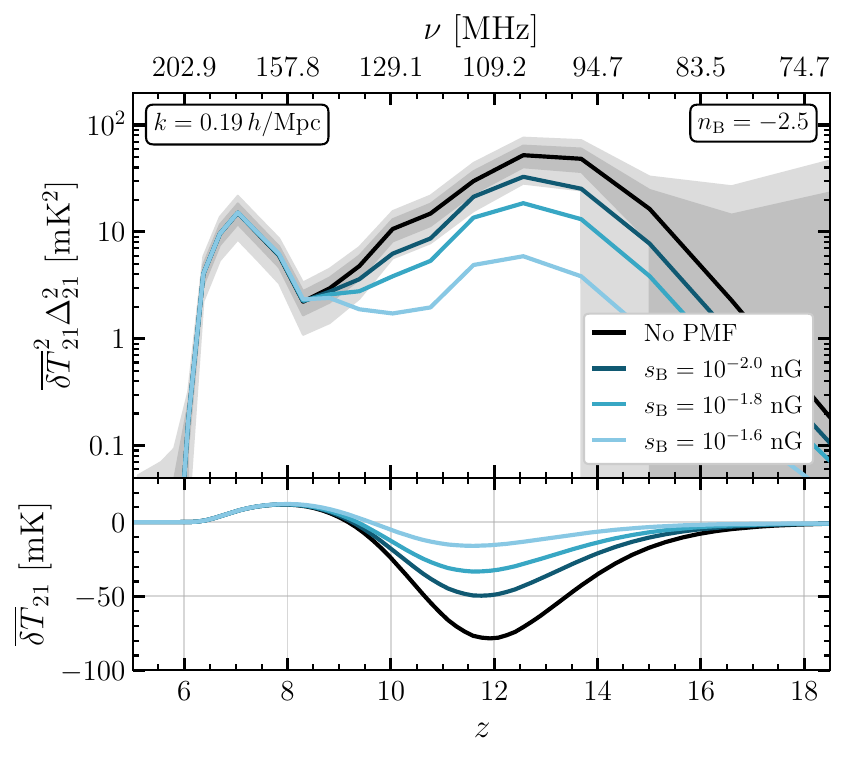}
    \caption{$21\,$cm power spectra at a fixed wavemode $k$ (\textit{upper panels}) and global signals (\textit{lower panels}) as a function of redshift for a fixed $n_{\rm B}=-2.5$ and different values of $s_{\rm B}$ (colored lines) as well as the prediction without PMFs (black line), which corresponds to the limit $s_{\rm B}\to0$ or $n_{\rm B}\to-3$. This plot was made using the fiducial astrophysical parameters listed in Tab.~\ref{tab:astro_params}.}
    \label{fig:PMF_21cm_power_spectra}
\end{figure}

 The $21\,$cm signal is observed in contrast with the expected $21\,$cm photon intensity from the CMB. It is characterized by   the differential brightness temperature \cite{Furlanetto:2006jb}
\begin{align}
    \delta T_{\rm 21}\approx 20\,\textrm{mK}\left(\frac{T_{\rm S}-T_{\textrm{CMB}}}{T_{\rm S}}\right)x_{\textrm{HI}}(1+\delta_{\rm b})\left(1+\frac{1}{H}\frac{\textrm{d}v_r}{\textrm{d}r}\right)^{-1}\sqrt{\frac{1+z}{10}\frac{0.15}{\Omega_{\rm m} h^2}}\frac{\Omega_{\rm b} h^2}{0.023}\,,
    \label{eq:21cmdiffTb}
\end{align}
where $x_{\textrm{HI}}$ is the neutral fraction of hydrogen, $\delta_{\rm b}$ is the baryon density contrast, $\textrm{d}v_r / \textrm{d}r$ is the velocity gradient along the line of sight, $T_{\rm CMB}$ is the CMB black-body temperature and the spin temperature $T_{\rm S}$ is a measure of the relative occupancy of the two hydrogen energy levels. The observable of telescope arrays like HERA is the quantity $\overline{\delta T}_{21}^2(z)\Delta^2_{21}(k,z)$, where $\overline{\delta T}_{21}(z)$ is the spatial average of $\delta T_{\rm 21}$ and $\Delta_{21}^2$ refers to the dimensionless power spectrum of perturbations, which is related to the two-point correlation function of perturbations in $\delta T_{21}$ through a Fourier transform. \repl{One of the largest challenges in the measurement of the cosmological $21\,$cm signal is contamination by foreground, which are often several orders of magnitude larger in amplitude of the signal itself. In this paper, we focus on foreground avoidance, focusing on Fourier modes where foregrounds are expected to be subdominant w.r.t. the cosmological signal.}

In Fig.~\ref{fig:PMF_21cm_power_spectra}, we illustrate in the top panel the $21\,$cm power spectrum as a function of redshift at a fixed scale $k=0.19\,h/$Mpc \repl{(expected to be relatively free from foregrounds)} while the bottom panel shows the corresponding sky-averaged signal. Both have been obtained using {\tt exo21cmFAST}. The black lines correspond to the predictions for the fiducial astrophysical parameters reported in the last column of Tab.~\ref{tab:astro_params}, while the dark (light) gray region around this curve inform on the 1$\sigma$ (2$\sigma$) sensitivity of the HERA telescope assuming $1000$ hours of observations, for more details see below. For the displayed redshift range and fiducial model, Lyman-$\alpha$ photons emitted by the first astrophysical sources become the dominant source of spin flip-transitions through the Wouthuysen-Field effect~\cite{Wouthuysen:1952, Field:1958} around $z\sim 16$, thereby coupling the spin temperature $T_{\rm S}$ to the kinetic gas temperature $T_{\rm k}<T_{\rm CMB}$. This leads to a 21$\,$cm signal seen in absorption w.r.t. the CMB. At $z\sim 12$, X-rays emitted by astrophysical sources start to significantly heat the IGM up to the point where $T_{\rm k}>T_{\rm CMB}$, and the signal is seen in emission w.r.t. the CMB from $z\gtrsim 9$. At $z\sim8$, the emission of ionizing UV photons becomes significant, inducing  reionization at $z\sim 5$ and therefore $\overline{ \delta T}_{21}\to 0$. As a  rule of thumb, epochs during which the global 21$\,$cm signal evolves most rapidly are typically associated with astrophysically induced processes (Ly-$\alpha$ coupling, X-ray heating, and reionization)  that operate highly inhomogeneously and  therefore tend to produce peaks in the 21 cm power spectrum. All of these key features could receive modifications from exotic physics. 

The blue curves of Fig.~\ref{fig:PMF_21cm_power_spectra} illustrate the effect of PMFs on the observables for a fixed $n_{\rm B}=-2.5$ and increasing values of the PMF amplitude $s_{\rm B}$ between $s_{\rm B}=10^{-2}\,\rm nG$ (darkest blue line) and  $s_{\rm B}=10^{-1.6}\,\rm nG$ (lightest blue line). For the astrophysical model  considered, the PMFs mainly affect the $21\,$cm signal at $z\gtrsim 9$, leading to a shallower absorption through in $\overline{\delta T}_{21}$ and a corresponding suppression of the $21\,$cm power spectrum amplitude. These are clear signatures of IGM heating from exotic physics well before heating from astrophysical sources becomes relevant. The heating reduces the difference between $T_{\rm S }\sim T_{\rm k}$ and $T_{\rm CMB}$, thereby reducing the amplitude of the signal in absorption (see Eq.~(\ref{eq:21cmdiffTb})) and leading to a more homogeneous heating of the IGM, which suppresses the spin temperature perturbations captured by $\Delta^2_{21}$, see also e.g.~\cite{Valdes:2007cu, Evoli:2014pva,  Lopez-Honorez:2013lcm,   Facchinetti:2023slb,  Mena:2019nhm,Sun:2023acy, Agius:2025xbj, Sun:2025ksr, Agius:2025nfz}. From Fig.~\ref{fig:PMF_21cm_power_spectra}, it is evident that for the exemplary case where $n_{\rm B}=-2.5$, PMFs with amplitudes as small as $s_{\rm B}=10^{-1.8}\,$nG are already expected to shift the $21\,$cm power spectrum at the limit of the $2\,\sigma$ sensitivity of HERA. As visible in Fig.~\ref{fig:constraints}, such suppressed PMF amplitudes  are well beyond the reach of the current cosmological probes. We can thus expect future HERA measurements to strongly improve our sensitivity to PMFs.

\subsection{Forecast methodology}
\label{sec:forecast}

As illustrated in Fig.~\ref{fig:PMF_21cm_power_spectra}, the dominant effect of PMFs on the $21\,$cm signal is a suppression of the high-redshift peak of the low-$k$ power spectrum. This behavior is characteristic of exotic heating of the intergalactic medium before astrophysical X-ray heating becomes dominant. As discussed in Sec.~\ref{sec:ADvsDT}, PMF heating in this redshift range is expected to be driven by ambipolar diffusion. To consistently account for degeneracies between $n_{\rm B}$, $s_{\rm B}$ and astrophysical parameters, we perform a range of Fisher forecasts, where we fix $n_{\rm B}$ in each forecast and thus obtain marginalized bounds on $s_{\rm B}$.
 
 Our treatment of the Fisher forecasts follows Refs.~\cite{Facchinetti:2023slb, Dandoy:2025twl}. The Fisher information matrix for a set of parameters $\{\theta_i\}$ is given by
\begin{align}
    F_{ij} \equiv -\mathbb{E}_{\mathcal{O}} \left[ \left. \frac{\partial^2 }{\partial \theta_i\, \partial \theta_j} \ln \mathcal{L}(\mathcal{O} \, |\, \theta)  \, \right| \, \theta \right]  \, .
\end{align}
Here, $\mathcal{L}$ is the likelihood of the observed data $\mathcal O$ given a model and $\mathbb{E}_{\mathcal{O}}$ denotes the associated expectation value. The minimum covariance of two parameters is then given by $\mathrm{cov}(\theta_i, \theta_j)=(F^{-1})_{ij}$ at the Cram\'er-Rao bound \cite{frechet1943extension, darmois1945limites, aitken1942xv}, and the minimum marginalized standard deviation on a single parameter $\theta_i$ therefore is $\sigma_i= [(F^{-1})_{ii}]^{1/2}$. Under the assumption of a Gaussian likelihood, the Fisher matrix elements can be computed as
\begin{equation}
  F_{ij}=\sum_{i_k}\sum_{i_z}\frac{1}{\sigma_{21}^2(k_{i_k},z_{i_z})}\frac{\partial\, \overline{\delta T}_{\rm 21}^2\Delta^2_{21}({i_k},{i_z})}{\partial\, \theta_i} \frac{\partial\, \overline{\delta T}_{\rm 21}^2\Delta^2_{21}({i_k},{i_z})}{\partial\, \theta_j}\,,
  \label{eq:fisher_matrix}
\end{equation}
where the sums run over all the wavenumber $k$ and redshift $z$ bins. We model the total measurement error in each $k$- and $z$-bin as
\begin{equation}
    \sigma_{21}^2(i_k,i_z) = \sigma_{\rm exp}^2(i_k,i_z) + \sigma_{\rm shot}^2(i_k,i_z) + \left[\varepsilon \times \overline{\delta T}_{\rm 21}^2\Delta^2_{21}(i_k,i_z)\right]^2\,,
    \label{eq:21cm_noise}
\end{equation}
where $\sigma_{\rm shot}$ accounts for the shot noise in each bin and we set $\varepsilon=0.2$ to account for modeling error. 
\repl{ $\sigma_{\rm exp}$ is meant to capture thermal noise contributions from the sky and the receiver, with $\sigma_{\rm exp}(i_k, i_z)\propto T_{\rm sys}^2(z_{i_z})$~\cite{Morales:2009gs, 2013AJ....145...65P}, where $T_{\rm sys}$ is the \emph{system temperature}. Here we assume:
\begin{equation}
    T_{\rm sys}(z) = 100~{\rm K} + 260~{\rm K}\left(\frac{\nu(z)}{150 ~{\rm MHz}}\right)^{-2.6} \sim   100~{\rm K} + 300~{\rm K}\left(\frac{1+z}{10}\right)^{2.6} \, ,
    \label{eq:Tsys}
\end{equation}
corresponding to the default  system temperature settings defined in  the python package \texttt{21cmSense}\footnote{\url{https://github.com/jpober/21cmSense}}~\cite{Pober:2012zz, Pober:2013jna} that we use to evaluate the observation noise, assuming 1000$\,$h of observations.
Furthermore, we restrict our analysis to $k>0.1$ Mpc$^{-1}$, as lower values of $k$ are expected to be foreground-dominated. However, different calibration uncertainties and residual foreground contamination could  degrade the sensitivity, increasing the effective noise level or reducing further the number of usable modes,  see e.g.~\cite{Liu:2019awk} for a discussion. A detailed assessment of these effects is beyond the scope of the present work.}

We perform the Fisher forecasts using \texttt{21cmCAST}\footnote{\url{https://github.com/gaetanfacchinetti/21cmCAST}}~\cite{Facchinetti:2023slb} and the simulations in \texttt{exo21cmFAST}. To obtain the derivatives in Eq.~(\ref{eq:fisher_matrix}), we perform simulations of $\overline{\delta T}_{\rm 21}^2\Delta^2_{21}$ first with no PMF contributions (i.e. $s_{\rm B}=0$) and then with a fixed tilt $n_{\rm B}$ and a small variation of the PMF amplitude,  $s_{\rm B}^{\rm vary}$, in order to take a finite-difference derivative. Ideally, the value of the derivative should be independent of the value of $s_{\rm B}^{\rm vary}$. In practice,  this is subject to numerical noise and artifacts. We therefore perform simulations over a log-range of values of $s_{\rm B}^{\rm vary}$ and report a forecast uncertainty range. Out of numerical necessity, we choose a linear prior on $s_\B/{\rm nG}\in[0,\infty)$.

\subsection{Results}
\label{sec:21cmresults}

 We adopt a fiducial astrophysical model consistent with our   MCMC results  of Sec.~\ref{sec:analys_and_results}, summarized in the right column of Tab.~\ref{tab:astro_params}.
The Fisher forecast constraint   at 95\% CL for HERA,  computed as described above, is projected  in the  $(n_\B, s_\B)$ plane of Fig.~\ref{fig:constraints}, excluding the  cross-hatched light red region. The numerical uncertainty on the forecast,  stemming from the Fisher methodology, is  shown as the red band around the red  line limiting the cross-hatched region.   If $s_\B^{\rm vary}$ is chosen too small, numerical noise dominates and $\sigma_{ s_\B}\propto s_\B^{\rm vary}$. In contrast, if $s_\B^{\rm vary}$ is too large, the response of $\overline{\delta T}_{\rm 21}^2\Delta^2_{21}$ to $s_{\rm B}^{\rm vary}$ becomes non-linear and the finite-difference derivative cannot be computed meaningfully. The edges of the red shaded area in Fig.~\ref{fig:constraints}  shows the boundaries of these two regimes, while the solid red line shows the mean value of $\sigma_{s_\B}$ inside the interval. 
For the fiducial astrophysical scenario considered here, the forecast $21\,$cm bound is roughly one order of magnitude stronger than the combined {\bf Planck-$\tau$} + {\bf UV LFs} + {\bf EoR} constraint. 
In the redshift range most relevant for HERA, ambipolar diffusion provides the dominant heating contribution. Both the existing constraints and the projected $21\,$cm sensitivity depend primarily on the strength of the Lorentz force.  This can again  be seen by comparing the trend of the 21$\,$cm bound to the dashed gray lines  in Fig.~\ref{fig:constraints} that indicate constant values of the spatial average of the Lorentz force.

\begin{figure}
    \centering
    \includegraphics[width=0.7\linewidth]{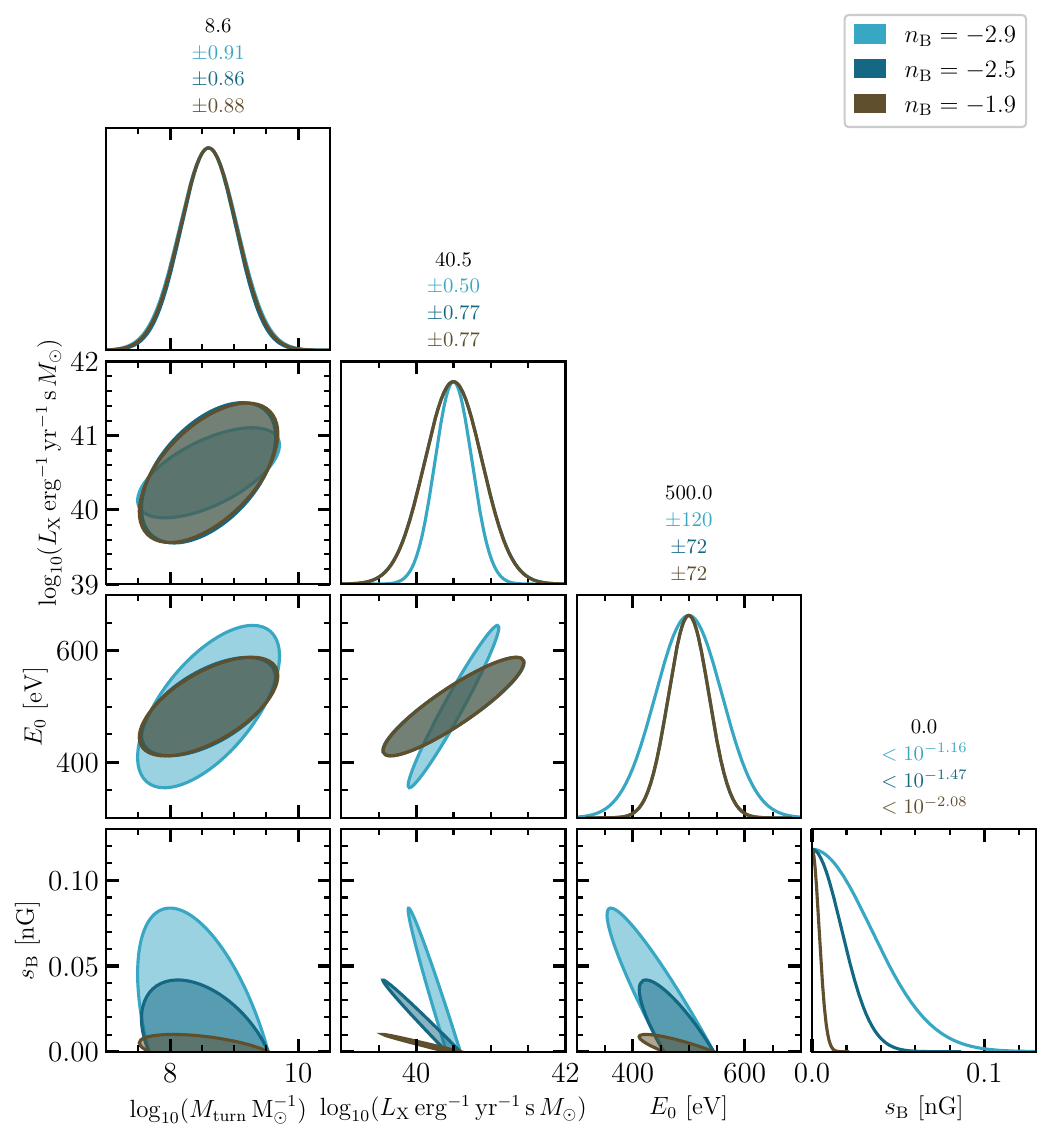}
    \caption{Triangle plot showing the 1D and 2D posteriors of $M_{\rm turn}$, $L_{\rm X}$, $E_0$ and $s_{\rm B}$ at $95\%$ CL for three different values of $n_{\rm B}$. Bounds on each parameter at $95\%$ as well are shown in colored text above the 1D posteriors. The fiducial value of each parameter is shown in black. The  posteriors in the $(M_{\rm turn}, L_{\rm X})$, $(M_{\rm turn}, E_0)$ and  $(L_{\rm X},E_0)$ planes are the same for $n_{B}=-1.9$ and $n_{B}=-2.5$ and thus almost exactly overlap each other.}
    \label{fig:triangle21}
\end{figure}

In Fig.~\ref{fig:triangle21} we show at 95\% CL 1D and 2D posteriors of a selection of parameters that could show some level of degeneracy with the amplitude of the PMF strength for three values of the PMF tilt going from $n_{\rm B}=-2.9$ (light blue) to $n_{\rm B}=-1.9$ (brown). Unsurprisingly, we find that $s_{\rm B}$, driving the strength of PMF heating, is strongly degenerate with $L_{\rm X}$, driving X-ray heating from stellar objects. Increasing values of $s_{\rm B}$ can be compensated with smaller values of $L_{\rm X}$ inducing a negative correlation between those two parameters. A negative correlation can also be observed between $s_{\rm B}$ and the  threshold energy for X-rays to escape galaxies,  $E_0$. This can be explained as follows. $L_{\rm X}$ is defined as the X-ray luminosity at energies above $E_0$, see Eq.~(\ref{eq:LX}). Lowering $E_0$ at a fixed value of $L_{\rm X}$ decreases the X-ray luminosity at the highest energies that is expected to heat the IGM homogeneously (as harder X-rays have a longer mean free path). This suppression in homogeneous IGM heating can be compensated by increasing the PMF heating, increasing their strength $s_{\rm B}$. Improving our knowledge of Cosmic Dawn galaxies, and in particular of their X-ray heating efficiency, could help to further improve constraints on exotic sources of heating.

\begin{figure}
    \centering
    \includegraphics[width=0.5\linewidth]{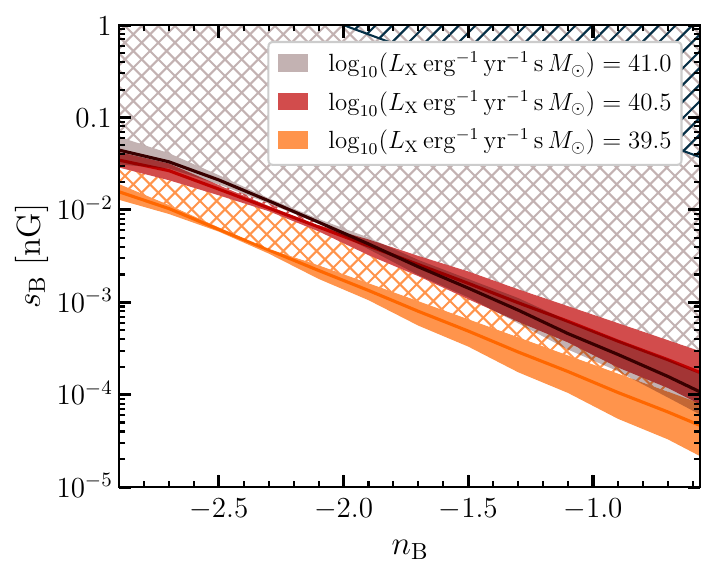}
    \caption{Forecast constraints on the PMF parameter space from HERA for three different fiducial values of $L_{\rm X}$. The width of the colored bands correspond to the numerical uncertainty of the Fisher forecasts, see Sec.~\ref{sec:forecast} and~\ref{sec:21cmresults}.}
    \label{fig:foreast_L_X_comparison}
\end{figure}

\repl{A large source of uncertainty in any forecasts for $21\,$cm cosmology is the relatively unknown astrophysics relevant to the description of the IGM evolution at the time of interest. As has been emphasized by previous works, certain values of astrophysical parameters could bias or even prevent us from improving on existing bounds on exotic sources of energy injection, see e.g.~\cite{Lopez-Honorez:2013lcm, Facchinetti:2023slb, Agius:2025xbj, Agius:2025nfz}. Here, we show how different choices of X-ray heating parameters can affect the forecast uncertainty.} Fig.~\ref{fig:foreast_L_X_comparison} shows the obtained uncertainty on $s_{\rm B}$ for three different values of $L_{\rm X}$, namely $\log_{10}(L_{\rm X}\,{\rm erg}^{-1}\,{\rm yr}^{-1}\,{\rm s}\,{\rm M}_\odot)=\{39.5,40.5,\,{\rm and}~41.0\}$. The lower value of $39.5$ corresponds to the lower bound on the X-ray amplitude from HERA data~\cite{HERA:2021noe} and gives rise to the $21\,$cm power spectrum with the largest amplitude, as it experiences the least suppression due to heating from astrophysical sources, which allows for a better sensitivity. We furthermore use the larger value of $41.0$ as a more pessimistic astrophysical scenario, as in e.g.~\cite{Facchinetti:2023slb, Agius:2025xbj}, because it leads to a comparative suppression of the power spectrum. In Fig.~\ref{fig:foreast_L_X_comparison}, we therefore project our obtained Fisher forecasts into the $(s_{\rm B}, n_{\rm B})$ plane, analogous to our summary plot of Fig.~\ref{fig:constraints},  keeping the fiducial values of all astrophysical parameters equal to the ones tabulated in the last column of Tab.~\ref{tab:astro_params}, except for $L_{\rm X}$. The orange hatched region shows the $21\,$cm sensitivity for $\log_{10}(L_{\rm X}\,{\rm erg}^{-1}\,{\rm yr}^{-1}\,{\rm s}\,{\rm M}_\odot)= 39.5$ while the gray hatched region  is obtained for $\log_{10}(L_{\rm X}\,{\rm erg}^{-1}\,{\rm yr}^{-1}\,{\rm s}\,{\rm M}_\odot)= 41.0$. Lower values of $L_{\rm X}$ lead to better constraints on the PMF strength $s_\B$, which would represent an improvement over current constraints by almost two orders of magnitude (see the blue region in Fig.~\ref{fig:constraints}). The degeneracy between $s_\B$ and $L_\X$ visible in Fig.~\ref{fig:triangle21} persists in all cases and provides a major obstacle to the ability to place stronger bounds on $s_\B$. \repl{Note that on general grounds, AD depends on the local ionization fraction and may therefore dissipate at different rates in ionized and neutral regions. This could, in principle, introduce additional spatial fluctuations in the magnetic field strength and heating rate around the Epoch of Reionization (EoR), potentially affecting the EoR peak differently when varying the PMF parameters. This may help in breaking the degeneracy between PMF and X-ray parameters. It is however beyond the scope of this paper to address this point in details, AD being assumed to mainly affect the ionized fraction in mainly neutral regions and the PMF energy injection being assumed to be homogeneous.}

\repl{We have also derived the forecast constraints that can be expected to be obtained with the upcoming \texttt{SKA-Low} $21\,$cm experiment~\cite{dewdney2009square} in its \texttt{AA4} (final) configuration. It is expected to improve on the HERA forecast projected in Fig.~\ref{fig:constraints} by approximately a factor $\sim 2$,  see Appendix~\ref{app:ska_forecast} for details.}

As discussed above, the dominant imprint on the 21$\,$cm signal arises from exotic heating of the IGM via ambipolar diffusion. Primordial magnetic fields are also expected to increase the halo abundance at primarily low halo masses, see Sec.~\ref{sec:HMF}. If this enhancement of the halo mass function happened at masses greater than $M_{\rm turn}$, it would lead to earlier and more significant star formation, which would in turn shift the features of the signal towards earlier times. However, we find that for the values of $s_{\rm B}$ to which the 21$\,$cm signal is expected to be sensitive to ($s_{\rm B}\lesssim0.05\,$nG for $n_{\rm B}\sim-2.9$), the enhancement of the HMF happens at masses that are too low to significantly affect the star formation rate. 
This can be seen comparing our fiducial choice of $M_{\rm turn}=10^{8.6}\,{\rm M}_\odot$ to the largest halo masses affected by the PMF boost of small scale structures, shown in Fig.~\ref{fig:hmf_mthresh_contour},  for the set of values of $(n_\B, s_\B)$ that lie at the lower limit of the cross-hatched regions of Fig.~\ref{fig:constraints}.   The region where $M_{\rm thresh}\sim M_{\rm turn}$ already lies within, or at the limit of, the area of the parameter space that HERA is expected to be sensitive to.  
 We have checked that, considering one single population of galaxies with a threshold mass for star-forming galaxies $M_{\rm turn}\gtrsim 10^6\, {\rm M}_\odot$, the main impact on the 21$\,$cm signal arises from heating and not from the PMF boost of density perturbations (and no degeneracy between $s_{\rm B}$ and $M_{\rm turn}$ appears).
This conclusion however likely depends on Cosmic Dawn galaxy population assumptions. Including minihalos which host molecular cooling galaxies, with threshold masses $\lesssim 10^6\,{\rm M}_\odot$, may allow the small-scale power enhancement to leave a detectable imprint. This will again depend on the astrophysical parameters associated to such an early star-forming galaxy population, and will  introduce a number of additional astrophysical parameters which may potentially be degenerate with $s_{\rm B}$ that one would have to take into account in a full analysis. Here, in order to stay consistent with our MCMC EoR+UV LF analysis of Sec.~\ref{sec:analys_and_results}, we do not account for minihalos. In the triangle plot of Fig.~\ref{fig:triangle21}, we see that $M_{\rm turn}$ and $s_{\rm B}$ are very mildly correlated (as compared to the X-ray parameters) and the 1D posterior of $M_{\rm turn}$ is essentially unaffected by the choice of $s_\B$. For completeness, a full triangle plot will all parameters considered is provided in appendix~\ref{app:triangle21}.

\repl{Also note that in our analysis, $\chi_B(z)$ is assumed to be independent of position and is computed from spatially averaged heating rates, see Eq.~(\ref{eq:chiBdot}). As reionization is an inhomogeneous process and  the AD heating rate scales as $x_e$, we can expect inhomogeneous PMF heating to source extra inhomogeneities potentially  leading to  an enhancement of the power in $\Delta_{21}^2$ at certain scales and redshifts. On the other hand, this could  also induce additional correlations between exotic heating sourced by AD and heating from stars hosted in galaxies. Even though a dedicated analysis, necessary to evaluate the impact of PMF inhomogeneous energy injection,  is beyond the scope of this work, such a study has been performed in~\cite{Sun:2023acy} in the context of dark matter energy injection from decays. Both the enhancement in power $\Delta_{21}^2$ and stronger degeneracies between astrophysics and exotic heating parameters were observed, which lead to projected sensitivities that were in the end similar to the homogenized treatment. Based on this, we expect our treatment of PMF energy injection at late times to provide a reasonably good approximation of the projected constraints from the $21\,$cm power spectrum.}

As mentioned in the introduction, gamma-ray observations provide a lower bound on the IGMF strength, while an upper bound is inferred from the absence of a redshift evolution in the rotation measures of extragalactic sources. For coherence lengths of $\mathcal O(\rm Mpc)$ scales, the lower limit on the magnetic field strength is of the order of $B \gtrsim 10^{-17}\,\mathrm{G}$~\cite{Fermi-LAT:2018jdy, Blunier:2025ddu}. This constraint is expected to be significantly improved through deep observations of the nearest blazars with the Cherenkov Telescope Array (CTA), potentially reaching sensitivities of up to $B \sim 10^{-11}\,\mathrm{G}$~\cite{Korochkin:2020pvg}.\footnote{Fermi-LAT and CTA estimates in~\cite{Korochkin:2020pvg} assume a Kolmogorov magnetic power spectrum with spectral index $n_\B=-11/3$.} On the other hand, the current upper bounds from rotation measures and CMB observations are at the level of $10^{-9}$--$10^{-10}\,\mathrm{G}$. A potential discovery of the sources of ultra-high-energy cosmic rays (UHECRs) may also help probe the IGMF in local voids of the large scale structure with strengths down to $\sim 10^{-10}$--$10^{-11}\,\mathrm{G}$ range~\cite{Neronov:2021xua, Durrer:2013pga}. If such a magnetic fields are of primordial (cosmic) origin, we have shown that upcoming 21$\,$cm cosmology experiments could improve upon current upper bounds by one to nearly two orders of magnitude, achieving sensitivities down to $\sim 10^{-10}$--$10^{-11}\,\mathrm{G}$. Consequently, 21$\,$cm observations could serve as a complementary probe to CMB and UHECR measurements, providing stringent upper bounds on the IGMF and helping to close the existing sensitivity gap in combination with future gamma-ray experiments.

\section{Conclusion}
\label{sec:concl}
In this work, we have revisited cosmology constraints on primordial magnetic fields (PMFs) by incorporating a more realistic treatment of their evolution and energy losses from recombination to the epoch of reionization. Unlike previous analyses that mostly  assumed a redshifting of the PMF energy density as $(1+z)^4$, we have explicitly modeled the decay of the magnetic energy density due to ambipolar diffusion and decaying turbulences, two mechanisms that are known to significantly affect the thermal and ionization history of the intergalactic medium (IGM). Our approach shows that PMFs can lose a substantial fraction of their energy before reionization, reducing their impact on cosmological observables as compared to earlier estimates.

For our analysis, we have implemented the PMF-induced heating and ionization effects in the recombination code {\tt HyRec}, which we use to model the evolution of the intergalactic medium (IGM) above redshift $z=35$, and in the semi-numerical framework {\tt exo21cmFAST}, which models the IGM evolution including the ionized fraction as well as star formation between $z=35$ the epoch of reionization. To overcome the computational cost of computing the ionized fraction in the large astrophysical parameter space considered, we employ the {\tt NNERO} emulator, which is trained on {\tt exo21cmFAST}, to predict both the ionization history and optical depth. This allows us to perform a MCMC analysis to combine multiple late-time probes: the optical depth to reionization from Planck, direct measurements of the ionized fraction at $z \sim 6$, and the UV luminosity functions (UV LFs) of galaxies.

Our results demonstrate that accounting for PMF energy losses significantly relaxes previous bounds on PMF strength. The suppression of the PMF amplitude between recombination and reionization reduces their imprint on both the ionization history and the matter power spectrum at small scales. Consequently, the constraints derived in this work are weaker by roughly an order of magnitude compared to analyses that neglected PMF decay. 

Furthermore, the inclusion of UV LFs as an additional probe represents an extra contribution of this study. While UV LFs primarily constrain astrophysical parameters related to star formation, they also provide complementary information on small-scale structure formation, thereby helping to break degeneracies between PMF and astrophysical effects. 

Furthermore, we have  performed Fisher forecasts to self-consistently evaluate the future sensitivity of the HERA telescope array, which probes the properties of the IGM as well as spatial variations therein through the 21$\,$cm signal which arises from Cosmic Dawn. Assuming one single population of galaxies with, we show that for $M_{\rm turn}> 10^6\, {\rm M}_\odot$, the strongest impact of PMF on the 21$\,$cm signal arises from the heating through ambipolar diffusion. We show that the latter effect depends on the astrophysical scenario considered and shows a strong degeneracy with the amount of X-ray heating from galaxies. We conclude that 21$\,$cm cosmology can improve on current bounds on $s_{\rm B}$ by one to two orders of magnitude, depending on the choice of fiducial astrophysical parameters. 

In the future, several further improvements could be made, including using a star formation prescription that takes into account magnetic fields and obtaining a more accurate estimate of the effect of PMF on small-scale structure formation using MHD simulations with MHD over a wide redshift and parameter space, such as done in~\cite{Ralegankar:2024arh}. It would also be interesting to account for the fact that the presence of PMF may lead to larger baryon fractions inside halos.

In summary, this study highlights importance of accounting for PMF energy losses when studying their imprint on cosmological observables. Our analysis relaxes previous bounds while introducing UV luminosity data as a complementary  constraint.  Furthermore, forthcoming $21\,$cm observations from experiments such as HERA and SKA will provide direct probes of the IGM during Cosmic Dawn and reionization, and in the view of our analysis, would  outperform the current best probes. This implies that 21$\,$cm observations could serve as a complementary probe to close the existing sensitivity gap to IGMF, if they are shown to be of primordial origin, in combination with future gamma-ray experiments.

\appendix

\section{Extra material}

\subsection{Matter power spectrum boost}
\label{app:PSboost}

\begin{figure}[t]
    \centering
    \includegraphics[width=0.59\linewidth]{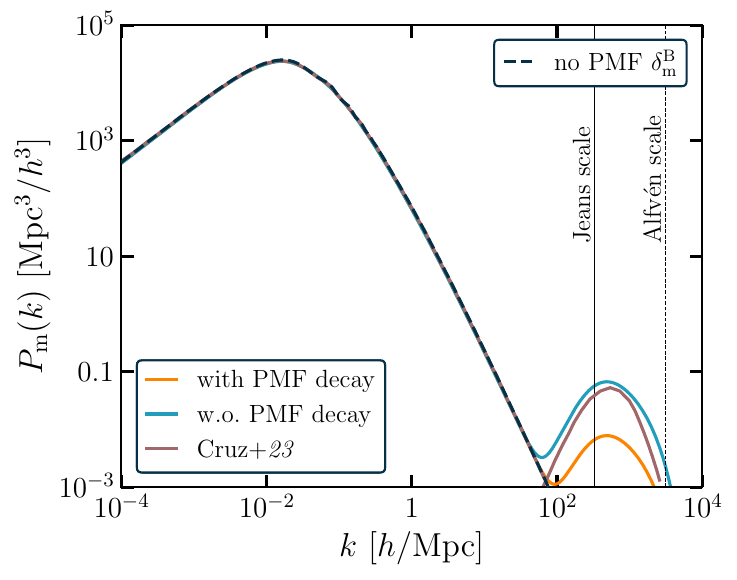}
    \caption{Linear matter power spectrum at $z=0$. The dark dashed line shows the standard $\Lambda$CDM spectrum while the colored peaks at large $k$ correspond to the PMF contribution ($s_{\rm B} = 10^{-2}\,$nG, $n_{\rm B} = -2$) in two different scenarios. We compare our result -- in yellow for decaying PMFs, in blue for stationary PMFs -- to the result of ref.~\cite{Cruz:2023rmo} in brown. The vertical lines show the two most important cutoff scales.}
    \label{fig:MPS_comp_Cruz}
\end{figure}

In Fig.~\ref{fig:MPS_comp_Cruz}, we compare our result for the linear matter power spectrum including the PMF contribution to that of \cite{Cruz:2023rmo}, see more details in Sec.~\ref{sec:growth}. We observe that the large-$k$ enhancement due to the presence of PMFs is similar in both cases if we consider that PMFs do not decay, up to a small difference, due to different conventions for the Alfv\'en scale. Considering the decay, the amplitude of the PMF-induced contribution to the matter power spectrum today is reduced by approximately an order of magnitude for the model considered.

\subsection{Full triangle plot of the 21$\,$cm cosmology Fisher forecast}
\label{app:triangle21}

\begin{figure}[t]
    \centering
    \includegraphics[width=0.99\linewidth]{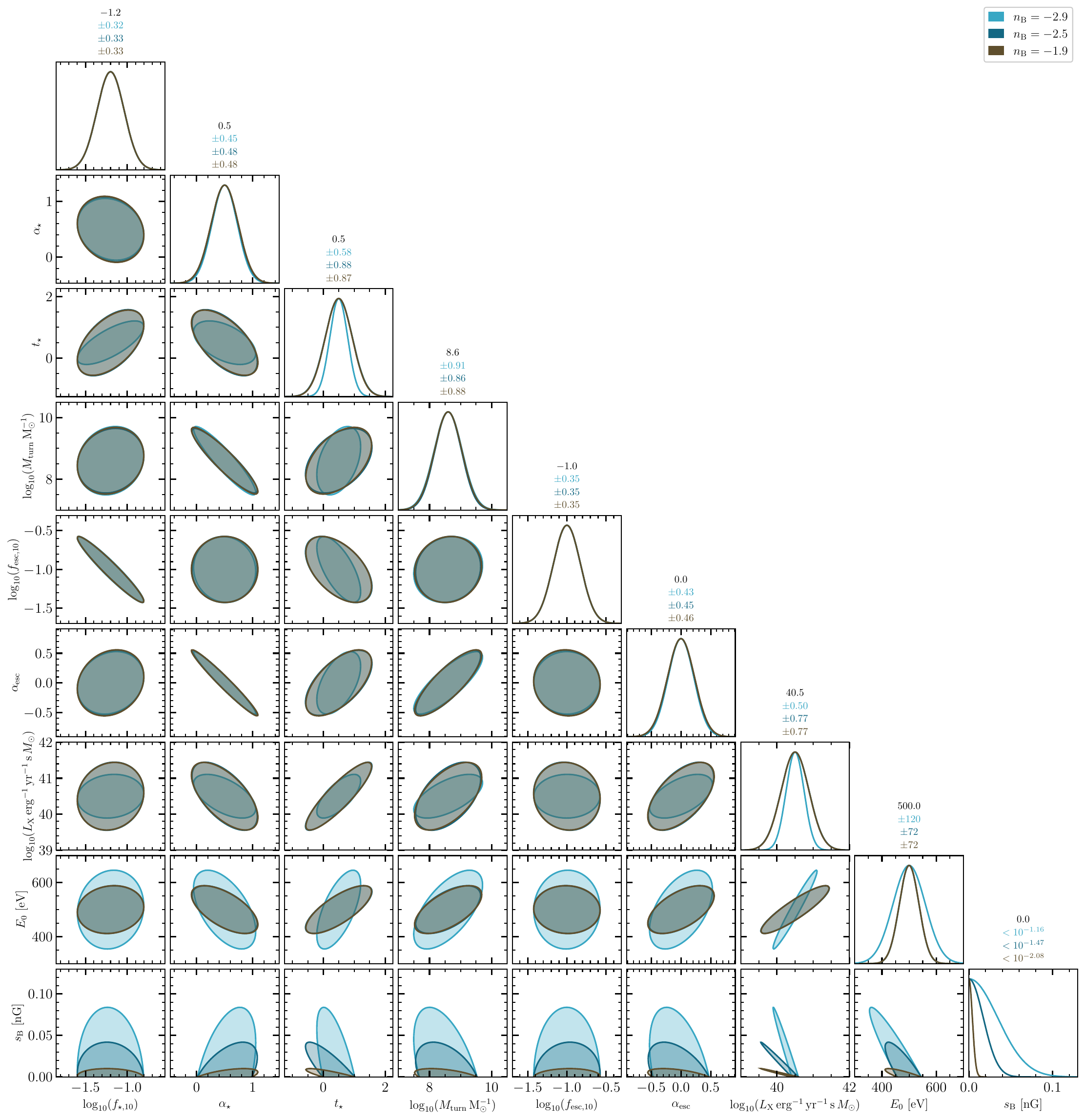}
    \caption{A full version of Fig.~\ref{fig:triangle21}, showing the degeneracies between all the parameters that we consider.}
    \label{fig:triangle21_large}
\end{figure}

In Fig.~\ref{fig:triangle21_large}, we show a full triangle plot resulting from the 21$\,$cm analysis performed in Sec.~\ref{sec:21cm} showing the degeneracies between all the parameters involved in our analysis for the fiducial astrophysical model of Tab.~\ref{tab:astro_params}.

\begin{figure}[h]
    \centering
    \includegraphics[width=0.5\linewidth]{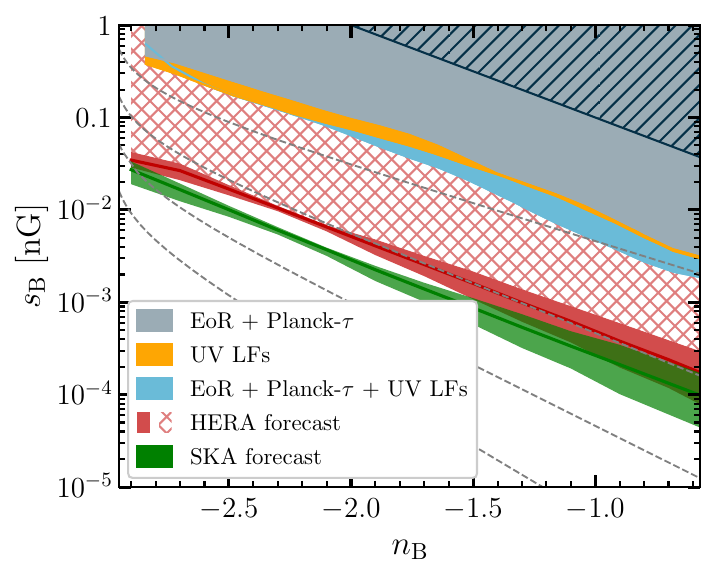}
    \caption{\repl{Same as Fig.~\ref{fig:constraints}, additionally showing forecast constraints for \texttt{SKA-AA4} (green).}}
    \label{fig:ska_forecast}
\end{figure}

\repl{\subsection{SKA AA4 Forecast}
\label{app:ska_forecast}

In addition to the forecast for HERA's ability to constrain PMFs with measurements of the cosmological $21\,$cm signal, we perform a forecast for the upcoming \texttt{SKA-Low} experiment in its \texttt{AA4} (final) configuration. This forecast follows exactly the prescription of our HERA forecast, except that we use \texttt{21cmSense} to calculate the experimental error of \texttt{SKA-AA4} instead of \texttt{HERA}. The resulting constraints are shown in green in Fig.~\ref{fig:ska_forecast}. We find an improvement in constraining power of approximately a factor of two.}

\acknowledgments
The authors would like to thank Y.~Qin for useful discussions and comments on the manuscript.  LLH, GF, and JRS have been 
 supported by the Fonds de la Recherche Scientifique F.R.S.-FNRS as a senior research
associate, a postdoctoral researcher and a doctoral research fellow, respectively.  The work of AK has been supported by the IISN
project No. 4.4501.18. LLH, GF, and JRS also acknowledge the support of the FNRS research
grant number J.0134.24 and of the ARC program of the Federation Wallonie-Bruxelles. All authors ackowledge the support of  the IISN
convention No. 4.4503.15. Computational resources have been provided by the Consortium
des Équipements de Calcul Intensif (CÉCI), funded by the Fonds de la Recherche Scientifique de Belgique (F.R.S.-FNRS) under Grant No. 2.5020.11 and by the Walloon Region
of Belgium. The authors are also members of BLU-ULB (Brussels Laboratory of the Universe, blu.ulb.be).

\bibliography{bibPMF}{} \bibliographystyle{unsrt}
\end{document}